\documentclass[english,prb,floatfix,superscriptaddress,twocolumn,showpacs,amsmath,amssymb,reprint]{revtex4-2}
\usepackage[T1]{fontenc}
\usepackage[latin9]{inputenc}
\usepackage{color}
\usepackage{textcomp}
\usepackage{amsmath}
\usepackage{amssymb}
\usepackage{graphicx}
\usepackage{wasysym}
\usepackage{gensymb}
\usepackage{amsmath}

\makeatletter

\providecommand{\tabularnewline}{\\}

\usepackage{bbold}
\usepackage{dsfont}
\PassOptionsToPackage{caption=false}{subfig} 
\usepackage{hyperref}
\hypersetup{
breaklinks=true,
colorlinks=true,
citecolor=blue,
linkcolor=blue,
filecolor=blue,
urlcolor=blue
}
\IfFileExists{lmodern.sty}{\usepackage{lmodern}}{}
\makeatother

\usepackage{babel}
\begin{document}
\title{Polarization as a tuning parameter for Floquet engineering:\\ magnetism in the honeycomb, square, and triangular Mott insulators}
\author{V. L. Quito}
\email{vquito@iastate.edu}
\affiliation{Department of Physics and Astronomy, Iowa State University, Ames,
Iowa 50011, USA}
\author{R. Flint}
\affiliation{Department of Physics and Astronomy, Iowa State University, Ames,
Iowa 50011, USA}
\date{\today}
\begin{abstract}
Magnetic exchange couplings can be tuned by coupling to periodic light, where the frequency and amplitude are typically varied: a process known as Floquet engineering.  The polarization of the light is also important, and in this paper, we show how different polarizations, including several types of unpolarized light, can tune the exchange couplings in distinct ways. Using unpolarized light, for example, it is possible to tune the material without breaking either time-reversal or any lattice symmetries. To illustrate these effects generically, we consider single-band Hubbard models at half-filling on the honeycomb, square and triangular lattices. We derive the effective Heisenberg spin models to fourth order in perturbation theory for arbitrary fixed polarizations, and several types of unpolarized light that preserve time-reversal and lattice symmetries.  Coupling these models to periodic light tunes first, second and third neighbor exchange couplings, as well as ring exchange terms on the square and triangular lattices.  Circularly polarized light induces chiral fields for the honeycomb and triangular lattices, which favors non-coplanar magnetism and potential chiral spin liquids.  We discuss how to maximize the enhancement of the couplings without inducing heating.
\end{abstract}
\maketitle

\section{Introduction \label{sec:Introduction}}

Floquet engineering, the process of using periodic drives to manipulate quantum matter has recently been applied, either experimentally or theoretically to a broad spectrum of materials, from graphene~\cite{Oka_2009_PRB,DemlerPRB2011,Foa_PRB_2014,Sie_Nature_2015,McIver_Nature_2020} and topological insulators~\cite{Lindner2011,Wang2013Science,Platero_PRL_2013,Neupert_PRL_2014,FietePRB2018} to frustrated magnetic insulators~\cite{sato_arxiv_2014,Mentink_NatPhys_2015,SatoOka_2016_PRL,Mentink_2017,ClaassenNatComm2017,Ishihara_PRB_2018,Refael_PRB_2019,Owerre_2019,Losada_PRB_2019,Millis_PRB_2019,Refael_PRB_2019,Mentink_2019_SciPost}. In particular, the use of lasers and ultrafast spectroscopy has proven to be a fruitful tuning knob for quantum matter, complementary to more conventional tools like pressure and magnetic field. Floquet engineering can even induce novel phases that do not exist in equilibrium (for a recent review, see Ref.~\onlinecite{Oka_review_2019}). The frequency and intensity of the laser light is typically varied to tune systems.  In this paper, we consider the effect of varying the \emph{polarization} of the laser light, focusing on magnetic insulators.  

Perfectly monochromatic light always has a fixed polarization that breaks a symmetry: either time-reversal/inversion in the case of circularly polarized light, or lattice symmetries for linear polarization; general elliptical polarizations break both symmetries. The symmetry breaking nature of polarization can be useful: for example, circularly polarized light can generate chiral fields in magnetic insulators that can drive chiral spin liquids~\cite{Oka_2009_PRB,ClaassenNatComm2017,Kitamura_PRB_2017}.  Linear polarization could be used to tune the anisotropy or dimensionality of a material~\cite{Halperin_PRX_2017,Refael_PRL_2018,Yuan_Optics_2018,Ozawa_NatureRev_2019,Dutt_Science_2020}.  For example, a lot of work has focused on anisotropic triangular lattice materials like Cs$_2$CuBr$_4$~\cite{Fortune2009} or organic materials like $\kappa$-(ET)$_{2}$Cu{[}N(CN)$_{2}${]}Cl~\cite{WelpPRL1992,MiyagawaPRL1995}, where two of the nearest neighbors have equal exchange couplings $J$, but the third has coupling $J'$.  The appropriate linearly polarized light could either reduce or enhance $J'/J$, allowing that axis of the phase diagram to be dynamically explored.  Two dimensional materials could be pushed towards one-dimensional physics fairly easily - for example, we find that reasonable fluences and frequencies can even change the sign of the nearest-neighbor coupling along a given direction, so that exchange couplings along a given direction could be tuned through zero.

Nevertheless, ideally, we would choose whether or not we broke a symmetry, and so unpolarized light is an appealing proposition. While perfectly monochromatic unpolarized light is a contradiction, it is possible to generate quasi-monochromatic unpolarized light either by varying the polarization vector slowly in time, or by passing the laser beam through an optical depolarizing element~\cite{burns1983Lightwave,mcguire90,hodgson2005laser}.  Even in the case where the polarization induces an additional time-scale, it is still possible to use the Floquet engineering formalism with additional averaging of the polarization at the end of the calculations~\cite{MukherjeePRB2018,QuitoFlintshort2020}.  In fact, different methods can be used to generate distinct types of unpolarized light distinguished by their higher order correlators~\cite{klyshko97}, which lead to different physical effects.

This proliferation of tuning abilities is particularly interesting for two-dimensional frustrated magnetic materials, which host a number of interesting phases and transitions~\cite{lacroix2011introduction}.  Many of these phases are hard to access, as it is rarely possible to experimentally tune across often multi-dimensional theoretical phase diagrams.  Spin liquids are of particular interest, as they typically break no symmetries, although chiral and nematic spin liquids can break time-reversal and rotational symmetries, respectively.  Spin liquids may have topological order with gapped spinons, or may host gapless spinon excitations~\cite{wen2004quantum}. Here we will show that changing the polarization of the periodic drive can substantially increase frustration, tuning through different parts of the phase space by modifying different magnetic exchange couplings in different ways.  It is even possible to change the relative sign of these exchange couplings simply by changing the polarization protocol. 

In this paper, we illustrate how very different the effects of different types of polarization can be using one of the simplest correlated electron models: the half-filled single-band Hubbard model.  Any two-dimensional spin liquid, either gapless or gapped~\cite{wen2004quantum}, could, in principle, be realized by our proposal. We can access both symmetric spin liquids, via unpolarized light, or chiral spin liquids, via circularly polarized light; our ability to access certain spin liquids within a single band Hubbard model depends on how much it is possible to enhance the appropriate further neighbor, chiral or ring exchange couplings.

 We consider three different lattices: honeycomb, square and triangular, and examine the driven magnetic exchange couplings to fourth order in perturbation theory over the full range of non-symmetry breaking unpolarized light. Some aspects of the triangular lattice case were explored in our previous work, Ref.~\cite{QuitoFlintshort2020}; we include it for completeness and to compare it to other lattices. On these lattices, we can examine further neighbor couplings, ring exchange terms, and chiral fields, and change the relative magnitudes, as well as the signs. Of course, most magnetic insulators are not really captured by a single-band Hubbard model, and are better described with magnetic exchange interactions coming from superexchange mediated by intermediate oxygens. However, superexchange is qualitatively similar to the processes considered here, but involving more orbitals and potentially higher order terms; we expect the results to be similarly tunable. Our description also neglects the phonons, which are expected to also contribute to the effective changes in hoppings, especially for strong electron-phonon couplings~\cite{Hubener_NanoLetters_2018,Shin_NatComm_2018,ChaudharyRefael_PRR_2020}.

In order for this method to be a practical experimental tuning method, it must not only be possible to tune the exchange couplings appropriately, but the material must also be able to quasi-equilibrate without significant heating, with sufficient time for the phase properties to be measured.  There are two issues here: are the time-scales for the different processes sufficiently far apart?  And is it possible to avoid heating, with high enough fluences to yield interesting results?  

There are four important time scales: first, the period of the Floquet pulse: $T=2\pi/\Omega$; the period of the polarization vector oscillation, $T_p$; the relaxation time of the spins $T_{rel} \sim \hbar/J \sim U/t_1^2$; and finally the overall length of the pulse, within which all tuning and measurements must be done.  The maximum enhancements of the exchange couplings are found for Floquet frequencies around $U$, which means the time-scales $T \sim 1/U \ll T_p \sim 10/U \ll T_{rel} \sim 100/U$ are all well separated for the small hoppings $t_1$ considered here.  If $\Omega$ is in the eV range, $T \sim 1$fs, and all time-scales fit easily within a pulse length on the order of a picosecond, which is moderately long for current experiments.

\begin{figure}
\includegraphics[width=1\columnwidth]{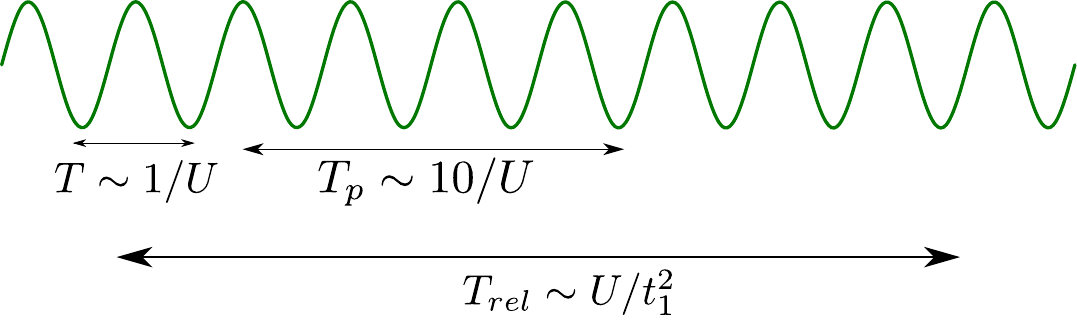}\caption{Time-scales involved in this problem. The laser frequency is $T=\frac{2\pi}{\Omega}$, while the polarization average can set a time scale $T_{p}$, etermined by the inverse of the Coulomb repulsion $U$. The spins equilibrate to their new exchange couplings within the relaxation time-scale, $T_{rel}$, given by $U$ and the nearest-neighbor hopping $t_{1}$. The measurements need to be made within the pulse after this relaxation is achieved.  \label{fig:The-time-scales}}
\end{figure}

The key to avoiding heating in a Mott insulator is to avoid photon frequencies that excite electrons across the Mott-Hubbard gap. As we consider relatively large fluences, it is necessary that no multi-photon processes can excite electron between the lower and upper Hubbard bands~\cite{Berges_PRL_2004,Abanin_mathem_2017,Mori_PRL_2016,RigolPRX2014,Lazarides_PRE_2014,SaitoAnnPhys2016}, which restricts not only the possible frequencies, but also the possible values of $t_1/U$ and thus the materials. The question of whether there is a transient regime in which the effective spin model is realized is not obvious a priori and has been addressed numerically on the kagom\'e lattice, with a positive answer~\cite{ClaassenNatComm2017}; these results should hold generally once the bandwidths are appropriately determined. Another possibility
to avoid heating that we do not address here is to couple the electronic system to a bath as, e.g phonons. In that case, the electrons can exchange energy with the heat bath. It may then be possible to develop transient non-equilibrium steady states~\cite{MitraPRB2014,MitraPRB2016,MiyashitaPRE2015,SeetharamPRX2015,ChamonPRB2015,AokiPRL2009}, with the non-equilibrium distribution functions emerging from the Floquet bands.

This paper is organized as follows. The choice of polarization, which
plays an important role in our later results, is introduced and discussed
in Sec.~\ref{sec:Polarization-choices}. In particular, we define
the types of unpolarized light that will be considered in this work. In Sec~\ref{sec:Model}
we define the model: the nearest-neighbor half-filled
Hubbard model in the presence of a Floquet field. In Sec~\ref{sec:General_results},
we give a brief review of the Floquet formalism and derive the effective
spin Hamiltonian from perturbation theory for arbitrary polarization.
The common features for the three lattices that we study are listed
in Sec~\ref{sec:General_results}, which includes the basic step
of the perturbation calculations, as well as the connections to experiments.
It is in Sec.~\ref{sec:Specific_lattices} that we show how the ratios
of exchange couplings are modified according to the polarization,
frequency, and fluence for each lattice. We present the results in
order of complexity, starting with the honeycomb in Secs~\ref{sec:Honeycomb_lattice},
then the square lattice in Sec.~\ref{sec:Square-lattice}, and finally,
the triangular one in Sec.~\ref{sec:Triangular lattice}. A summary
of our results and possible extensions are listed in Sec.~\ref{sec:Conclusions}.

\section{Polarization choices and averaging~\label{sec:Polarization-choices}}

In this section, we introduce our notation for treating different polarizations of light and discuss the different types of unpolarized light that may be generated.  Ultimately, we will compute properties for arbitrary polarizations and average over different polarization distributions to find the unpolarized results.  Unpolarized light may be generated from polarized light either using an optical depolarizer that spatially disorders the polarization, or by combining two laser beams with orthogonal polarizations and slightly detuned frequencies that cause the polarization vector to vary in time, with period $T_p$.  As long as this time-scale is sufficiently large compared to the Floquet period, $T= 2\pi/\Omega$, averaging over the polarization distribution function should give correct results~\cite{QuitoFlintshort2020}.

Let us assume that the propagation vector of the light is normal to the two-dimensional lattice and that the light is monochromatic, with frequency $\Omega$. The electric field is
\begin{equation}
\boldsymbol{E}\left(t\right)=\text{Re}\left[\boldsymbol{E}_{0}e^{-i\Omega t}\right],\label{eq:electric_field}
\end{equation}
with $\boldsymbol{E}_{0}$ independent of time. We can parameterize $\boldsymbol{E}_{0}$ for an arbitrary polarization, using either circular or linear polarization bases.
The calculation is substantially simpler if done with the circular basis,
$\boldsymbol{E}_{0}=E_{x}\hat{\epsilon}_{1}+E_{y}\hat{\epsilon}_{2}=E_{+}\hat{\epsilon}_{+}+E_{-}\hat{\epsilon}_{-}$,
with $\hat{\epsilon}_{\pm}=\frac{1}{\sqrt{2}}\left(\hat{\epsilon}_{1}\pm i\hat{\epsilon}_{2}\right)$.
In this basis, a generic polarization may be written as,
\begin{align}
E_{+} & =\sqrt{I}\sin\left(-\chi-\pi/4\right)e^{-i\left(\psi-\pi/2\right)},\cr
E_{-} & =\sqrt{I}\cos\left(-\chi-\pi/4\right)e^{i\left(\psi-\pi/2\right)}.\label{eq:Eminus}
\end{align}
This choice might appear unnecessarily complicated, but
proves convenient for our calculations, where the polarization can be captured by the two angles, which describe a sphere whose radius is $I$. The total intensity of the electromagnetic field is $I=\left|E_{+}\right|^{2}+\left|E_{-}\right|^{2}$. 

\begin{figure}
\includegraphics[width=1\columnwidth]{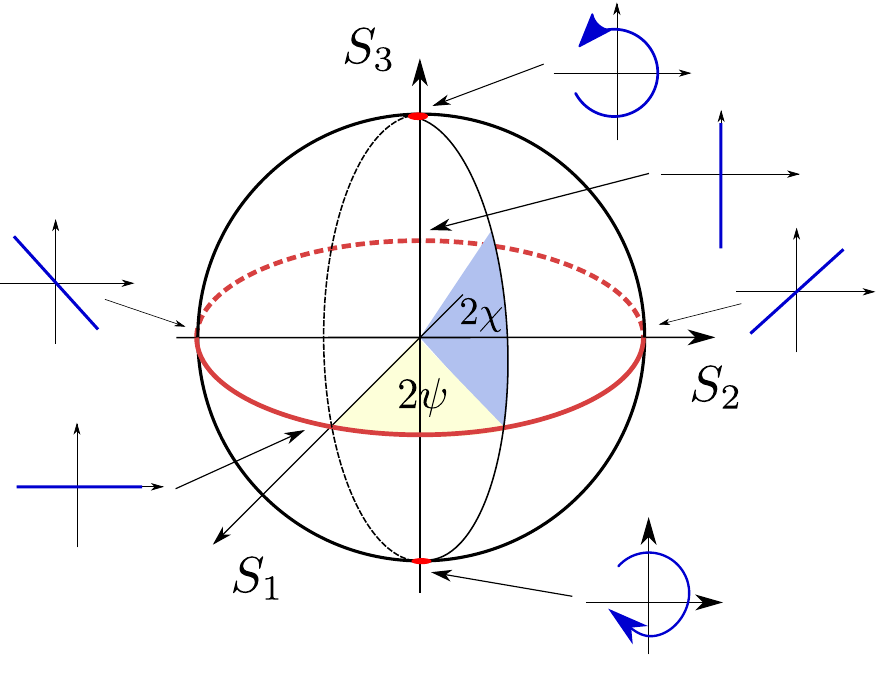}

\caption{Poincar\'e sphere, plotted in terms of the Stokes parameters $S_{1}$, $S_{2}$
and $S_{3}$, defined in (\ref{eq:stokes}).
A fixed point on the surface of the sphere defines light with a fixed
polarization. Circularly polarized right is located at the north and
south poles of the sphere. In this work, we also consider light configurations
with varying polarizations, that is, unpolarized light, of different
kinds. Unpolarized light obtained by averaging over linear polarization
(type II Glauber light) is represented at the red equatorial line.
Another type involves the average over the entire sphere (amplitude-stabilized
unpolarized light). \label{fig:Poincare-sphere}}
\end{figure}

A fixed polarization of light may be characterized using three quantities called Stokes parameters~\cite{bornwolf_book}, which are
\begin{align}
S_{1} & =2\text{Re}\left(E_{+}E_{-}^{*}\right)=I\cos2\chi\cos2\psi\cr
S_{2} & =2\text{Im}\left(E_{-}E_{+}^{*}\right)=I\cos2\chi\sin2\psi\cr
S_{3} & =\left|E_{+}\right|^{2}-\left|E_{-}\right|^{2}=I\sin2\chi.\label{eq:stokes}
\end{align}
These Stokes parameters span the surface of the Poincar\'e sphere, in terms of the angles $\left(2\chi,2\psi\right)$ and radius $I$. The Poincar\'e sphere is shown in Fig.~\ref{fig:Poincare-sphere}.
Some familiar cases can be recovered from Eq.~(\ref{eq:Eminus}).
Circular polarization is found by setting either $E_{+}$ or $E_{-}$
to zero, with $\chi = \pm \pi/4$. For these values, $S_{1}=S_{2}=0$ and $S_{3}$ has its maximum absolute value.
These are the north and south poles of the Poincar\'e sphere.
At $\chi=0$, the light is linearly polarized with $S_{3}=0$: the equator
of the sphere. Relative angles with the $S_{1}$ and $S_{2}$ axes determine
the polarization angle, $\psi \in (0,\pi)$. A schematic representation of different polarizations
as function of $\chi$ and $\psi$ is shown in Fig~\ref{fig:Poincare-sphere}.
For other generic values of $\chi$ and $\psi$, the light has elliptical
polarization, with all $S_i$ non-zero. 

\subsection{Partially polarized and unpolarized light}

While all perfectly monochromatic light is fully polarized, unpolarized and nearly monochromatic light may be created by allowing the polarization vector, $\vec{E}_0$ to slowly vary over the Poincar\'e sphere, with a characteristic time, $T_p = 2\pi/\Omega_p$ that is assumed to be much larger than the period, $T = 2\pi/\Omega$, such that the time average of the Stokes parameters are all zero, $\langle S_i \rangle = 0$~\cite{bornwolf_book,PiqueroOptics2018,ColasLight2015,BeckleyOptExp10,ortega17,shevchenko17,zhu15,shevchenko19,hannonen19}.  In this case, the light is quasi-monochromatic and unpolarized.  Partially polarized light may also be generated by allowing nonzero $\langle S_i \rangle$, such that $\sqrt{\sum_{i=1}^3 \langle S_i\rangle^2} < I$. Alternately, unpolarized light may be created by passing a fully polarized beam through an optical depolarizing element that makes the polarization vary rapidly over the spatial extent of the beam, such that the spatial average of the Stokes parameters are all zero, $\langle S_i \rangle = 0$~\cite{burns1983Lightwave,mcguire90,hodgson2005laser}. Technically these are called ``pseudo''-depolarizers as the resulting polarization is not random, however randomness is not required for our purposes.

Different polarization protocols or depolarizers create different types of unpolarized light that are differentiated by higher-order correlators of the Stokes parameters~\cite{klyshko97}, $\langle S_i S_j \rangle$, $\langle S_i S_j S_k S_l\rangle$, etc.  As our goal is to preserve the lattice and time-reversal symmetries, these higher-order correlators must also not break symmetries, which imposes restrictions on the allowed polarization distributions.  In general, we can specify a distribution, $f(I,\chi,\psi)$ that varies not only the angles, $\chi$ and $\psi$, but also, in principle, the intensity of the light.  Most of the time, we will fix the intensity, $\delta(I-I_0)$, as for laser light and consider just the angular distribution, $f(\chi,\psi)$.  However, the intensity variation is required to treat ``natural'' or thermal light.~\cite{bornwolf_book}

The higher-order correlators can be treated most straightforwardly by considering the spherical multipoles, $S_{lm}$ of the polarization distribution on the Poincar\'e sphere, $f(\chi,\psi)$, where
\begin{equation}
    S_{lm} = \!\!\int_{-\pi/4}^{\pi/4}\!\!\!d\chi\! \int_0^\pi\!\! d\phi \cos 2\chi f(\chi,\psi) Y_{lm}\left(\frac{\pi}{2}-2\chi,2\psi\right).
\end{equation}
Note that here we neglect the potential time dependence of time-dependent polarization protocols, where the path sampling the Poincar\'{e} sphere is traversed in a particular direction; this time-reversal symmetry breaking can be made arbitrarily small for $T_p >> T$~\cite{QuitoFlintshort2020}.

The magnetic exchange couplings are sensitive to these higher order correlators, and so are different for different types of unpolarized light.  There are two main classes of polarization distributions, type I and type II.  Type I is the most restrictive and samples the entire Poincar\'e sphere uniformly: $f(\chi, \psi) = 1$, which means only $S_{00}$ is nonzero, while type II light must be invariant under rotations, which restores lattice symmetries, and has zero net chirality, which restores time-reversal and inversion symmetries~\cite{LehnerPRA1996}, with only $S_{2n,0}$ nonzero.  It is generally important to check that the symmetry breaking multiples for a given depolarizer/polarization protocol vanish, as some depolarizers will sample the entire Poincar\'e sphere, but do so unevenly - e.g. - the dual Babinet compensator depolarizer~\cite{mcguire90} breaks four-fold lattice symmetries.

Type I light with a fixed amplitude, $I = I_0$ is known as amplitude-stabilized unpolarized light (ASUL)~\cite{bornwolf_book},
\begin{equation}
f\left(I,\chi,\psi\right)=\delta\left(I-I_{0}\right).\label{eq:f_ASUL}
\end{equation}
Natural light is the most familiar type of unpolarized light, and it is a particular kind of type I light where the intensity also varies exponentially, $f\left(I,\chi,\psi\right)=\frac{2}{I_{0}}\exp\left(-2I/I_{0}\right)$~\cite{goodman2015statistical}. In all of our work, due to the normalization, both kinds of type I light, ASUL and natural light give identical exchange couplings, although natural light is far from quasi-monochromatic and so not particularly relevant here.  In what follows, when we discuss type I light, we mean ASUL.  It may be generated via different techniques like a
coaxial superposition of modes of orthogonal polarizations~\cite{BeckleyOptExp10}
and, more recently, by shining a uniformly polarized beam into a uniaxial
crystal~\cite{PiqueroOptics2018}.

Type II light only allows $S_{2n,0}$ moments to be nonzero.  The most natural type II light is type II ``Glauber'' light, which samples all linear polarizations equally, with $\chi =0$.  Type II Glauber light can be generated by a linear combination of LCP and RCP light with slightly detuned frequencies~\cite{QuitoFlintshort2020} or by using a Cornu depolarizer~\cite{mcguire90}, for example.  More generically, we can construct type II light by equally sampling both $\pm \chi_0$,
\begin{equation}
f\left(I,\psi,\chi\right)=\frac{1}{2}\delta\left(I-I_{0}\right)\left[\delta\left(\chi-\chi_{0}\right)+\delta\left(\chi+\chi_{0}\right)\right].\label{eq:typeII_avg}
\end{equation}
All linear polarizations ($\psi$) are given equal weight, maintaining the lattice symmetry and forcing $\langle S_1\rangle = \langle S_2 \rangle = 0$. The absence of circular polarization, $\left\langle S_3\right\rangle =0$ is achieved by including both $\pm\chi_{0}$.  Type II Glauber light is the $\chi_0 = 0$ case. In Ref.~\onlinecite{ColasLight2015}, it has been demonstrated that Rabi oscillations of superimposed delayed fields, can cause the polarization vector to precess, covering circles on the surface of the Poincar\'e sphere. The precession period $T_{p}$
can be controlled, in order to keep $T_{p}\gg T$ and thus the quasi-monochromatic character of the beam. These generic type II distributions are somewhat artificial. However, they span the space of all type II light, and so we consider the full range of $\chi_0$'s to capture all possible type II distributions.  

Using the notation $J_{i\ldots}\left(I,\chi,\psi\right)$ to denote a generic
coupling calculated for a particular polarization and intensity, the
average is performed over a polarization distribution $f\left(I,\chi,\psi\right)$
according to
\begin{align}
\left\langle J_{i\ldots}\right\rangle =\frac{\int dV\left[2I\cos\left(2\chi\right)\right]\,f\left(I,\chi,\psi\right)J_{i\ldots}\left(I,\chi,\psi\right)}{\int dV\,\left[2I\cos\left(2\chi\right)\right]\,\,f\left(I,\chi,\psi\right)}\label{eq:avg_J-1}
\end{align}
with $\int dV=\int dI\,\int_{-\pi/4}^{\pi/4}d\chi\,\int_{0}^{\pi}d\psi$ the volume element of the Poincar\'e sphere. In the following sections, we will derive expressions for arbitrary polarization, $J_{i\ldots}\left(I,\chi,\psi\right)$ and then average using Eq. (\ref{eq:avg_J-1}) with different polarization distribution functions.

\section{Floquet-Hubbard model~\label{sec:Model}}

To illustrate the effect of different polarization protocols, we examine the single-band Floquet-Hubbard model with nearest-neighbor hopping on three different lattices.  While this model is certainly an oversimplification for most materials, it is the simplest in which the combination of interactions and the Floquet potential lead to non-trivial results and can illustrate effects that will apply much more generally.  In this section, we introduce the model, the basics of Floquet theory, and discuss the full Floquet-Hubbard Hamiltonian for arbitrary light polarization. 

\subsection{Time-independent model~\label{subsec:Hubbard-model}}

While the single band Hubbard model is familiar, we use it to introduce our notation.  We separate the Hamiltonian into a hopping term, $\mathcal{V}$, with hopping parameter $t_1$ (as $t$ is reserved for time), and an interaction term, with Hubbard interaction $U$,
\begin{align}
\mathcal{H}_{0} & =\mathcal{V}+\mathcal{H}_{\text{int}}\cr
\mathcal{V} & =-t_{1}\sum_{\left\langle i,j\right\rangle }c_{i}^{\dagger}c_{j}-\mu\sum_{i}c_{i}^{\dagger}c_{i}\cr
\mathcal{H}_{\text{int}} & =U\sum_{i}n_{i\uparrow}n_{i\downarrow}\label{eq:H-int}
\end{align}
We consider nearest-neighbor hopping with the hopping directions labeled, $\boldsymbol{\delta}_{l}=\left(\cos\phi_{l},\sin\phi_{l}\right)$.  On the honeycomb lattices,
\begin{align}
\boldsymbol{\delta}_{1} & =\left(\frac{1}{2},\frac{\sqrt{3}}{2}\right),\,\boldsymbol{\boldsymbol{\delta}}_{2}=\left(1,0\right),\,\boldsymbol{\boldsymbol{\delta}}_{3}=\left(\frac{1}{2},-\frac{\sqrt{3}}{2}\right),\label{eq:delta-NN-triangular-honeycomb}
\end{align}
while on the triangular lattice we have six hoppings, $\pm\boldsymbol{\delta}_{l}$,
and on the square lattice we have four ($\pm \boldsymbol{\delta}_{l}$), with
\begin{equation}
\boldsymbol{\delta}_{1}=\left(0,1\right),\,\boldsymbol{\boldsymbol{\delta}}_{2}=\left(1,0\right).\label{eq:delta-NN-square}
\end{equation}
As we are interested in magnetic states, we shall restrict ourselves to the half-filled Hubbard model, with $\mu$ adjusted to fix the filling for a given lattice and $U\gg t_{1}$. For intermediate values of $U$, the metal-to-insulator transition in a driven triangular lattice has been addressed in Ref.~\onlinecite{jana2019PRB}. 

\subsection{Basics of Floquet theory~\label{subsec:Basics-of-Floquet}}

Now, we briefly review the basics of Floquet theory, which allows us to handle time-periodic Hamiltonians.  We begin with the Schr\"odinger equation
\begin{equation}
i\partial_{t}\left|\psi\left(t\right)\right\rangle =H(t)\left|\psi\left(t\right)\right\rangle .
\end{equation}
If a generic state of the time-independent problem is $\left|\varphi_{a}\right\rangle $,
the Floquet theorem states that the generic eigenstates of the periodic
Hamiltonian $H(t)=H(t+T)$ are~\cite{mahan2010book}

\begin{equation}
\left|\varphi_{a}\left(t\right)\right\rangle =e^{-iE_{a}t}\sum_{n}e^{i\Omega nt}\left|\varphi_{a}^{n}\right\rangle ,
\end{equation}
analogously to the Bloch theory for spatially periodic Hamitonians\footnote{Sometimes Bloch's theorem is referred to as an application of the Floquet theorem, and the term Floquet theorem englobes any periodic
potential.}.  Here, $\Omega=\frac{2\pi}{T}$, and the effect of the time-periodic potential is incorporated via the extra discrete degree of freedom, $n$.  We can insert these states into the Schr\"odinger equation and average over a whole period, $T$ to find an equation for the states 
$\left|\varphi_{a}^{n}\right\rangle $ and quasi-energies $E_{a}$,~\cite{Shirley_PR_1965,Sambe_PRA_1973,mahan2010book,AokiPRB2016}
\begin{equation}
\left(E_{a}-\Omega m\right)\left|\varphi_{a}^{m}\right\rangle =\!\sum_{n}\!\left(\!\frac{1}{T}\!\int_{0}^{T}\!\!\!dte^{-i\Omega mt}H(t)e^{i\Omega nt}\left|\varphi_{a}^{n}\right\rangle\! \right).
\end{equation}
Since $H(t)$ is periodic in time, it admits the Fourier expansion $H(t)=\sum_{m}H_{m}e^{i\Omega mt}$,
with coefficients

\begin{equation}
H_{m}=\frac{1}{T}\int_{0}^{T}dte^{-im\Omega t^{\prime}}H(t^{\prime}).\label{eq:Floquet-components}
\end{equation}
By using Eq.~(\ref{eq:Floquet-components}), the now time-independent
Schr\"odinger equation for the states $\left|\varphi_{a}^{m}\right\rangle $
reads

\begin{align}
\sum_{n}\left(H_{m-n}+\Omega m\delta_{mn}\right)\left|\varphi_{a}^{n}\right\rangle =E_{a}\left|\varphi_{a}^{m}\right\rangle. \label{eq:time-ind-schro}
\end{align}
We can then work with this effective Hamiltonian just as we would work with the original time-independent Hamiltonian. In the next subsection, we derive the components $H_{m}$ for the Hubbard model in an external potential. 

\subsection{Floquet Hamiltonian~\label{subsec:Floquet_Hamiltonian}}

In this work, we consider light normally incident on the material, with the direction of polarization in the plane of the lattice.  We incorporate this field via a Peierls substitution with the vector potential $\mathbf{A}(t)$, and obtain the time-dependent hopping Hamiltonian,

\begin{align}
\mathcal{V}\left(t\right) & =-t_{1}\sum_{\left\langle i,j\right\rangle }\exp\left(-i\int_{\boldsymbol{R}_{i}}^{\boldsymbol{R}_{j}}\boldsymbol{A}\left(t\right)\cdot d\boldsymbol{r}\right)c_{i}^{\dagger}c_{j}\nonumber \\
 & =-t_{1}\sum_{i,\boldsymbol{\delta}_{i}}\exp\left(-i\boldsymbol{A}\left(t\right)\cdot\boldsymbol{\delta}_{i}\right)c_{i}^{\dagger}c_{i+\boldsymbol{\delta}_{i}}.\label{eq:Hhopt}
\end{align}
The Floquet coefficients, $\mathcal{V}_m$ are found by integrating
(\ref{eq:Hhopt}) over a full cycle, where for simplicity we define $\theta=\Omega t$,

\begin{align}
\mathcal{V}_{m} & =-t_{1}\sum_{j}\frac{1}{2\pi}\int_{0}^{2\pi}d\theta e^{-im\theta}e^{-i\boldsymbol{\delta}_{i}\cdot\boldsymbol{A}\left(\theta\right)}c_{i}^{\dagger}c_{i+\boldsymbol{\delta}_{i}}\label{eq:H-hop-before-integration}\cr
 & =-\sum_{i}t_{i,i+\boldsymbol{\delta}_{i}}^{\left(m\right)}c_{i}^{\dagger}c_{i+\boldsymbol{\delta}_{i}},
\end{align}
where $t_{i,i+\boldsymbol{\delta}_{i}}^{\left(m\right)}$ are our new hopping terms between both sites and Floquet sectors,
\begin{equation}
t_{i,i+\boldsymbol{\delta}_{i}}^{\left(m\right)}=t_{1}\frac{1}{2\pi}\int_{0}^{2\pi}d\theta e^{-im\theta}e^{-i\boldsymbol{\delta}_{i}\cdot\boldsymbol{A}\left(\theta\right)}.\label{eq:t_m}
\end{equation}
Our Hilbert space now contains both the electronic Fock space and the discrete set of Floquet sectors, giving the full Hamiltonian,

\begin{align}
\mathcal{H} = &\sum_{m,n}\mathcal{V}_{n-m}\left|m\right\rangle \left\langle n\right|+\mathcal{I}_{\,\text{Fock}}\otimes\sum_{m}m\,\Omega\left|m\right\rangle \left\langle m\right|+\nonumber \\
 & +\mathcal{H}_{\text{int}}\otimes\mathcal{I}_{\,\text{Floquet}}\label{eq:Hamiltonian-honeycomb-Floquet}
\end{align}
It is instructive to represent this Hamiltonian as a matrix in the Floquet space,
\begin{equation}
\!\!\mathcal{H}=\left[\begin{array}{ccc}
\ddots & \vdots\\
\mathcal{V}_{0}+\mathcal{H}_{\text{int}}-\Omega & \mathcal{V}_{1} & \mathcal{V}_{2}\\
\mathcal{V}_{-1} & \mathcal{V}_{0}+\mathcal{H}_{\text{int}} & \mathcal{V}_{1}\\
\mathcal{V}_{-2} & \mathcal{V}_{-1} & \mathcal{V}_{0}+\mathcal{H}_{\text{int}}+\Omega\\
 & \vdots & \ddots
\end{array}\right].\label{eq:Hamiltonian_coupled_light}
\end{equation}

\begin{figure}
\includegraphics[width=0.75\columnwidth]{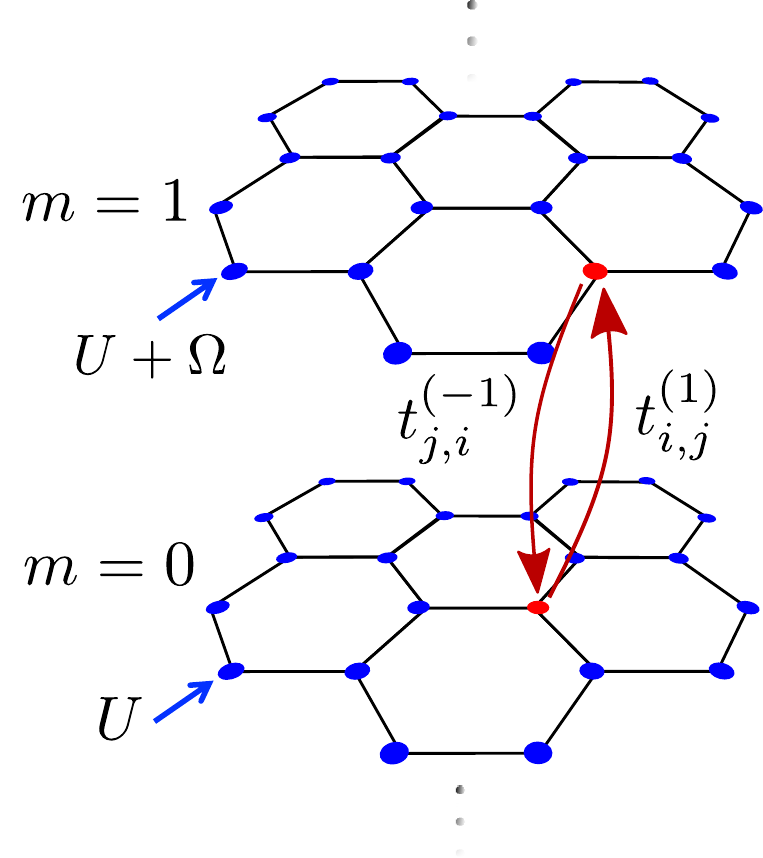}

\caption{Schematic representation of the Hubbard model on the honeycomb lattice
coupled to periodic light, following from Eq.~(\ref{eq:Hamiltonian_coupled_light}).
The Floquet index $m$ labels different copies of the honeycomb lattice, and
the hoppings are now between nearest-neighbor sites and Floquet sectors
(red dots). The on-site repulsion of the $m$-th copy is changed to
$U+m\Omega$.\label{fig:Floquet-sectors}}
\end{figure}

Effectively, an infinite number of copies of the lattice is created, each labeled by an integer $m$, as shown in Fig.~\ref{fig:Floquet-sectors}. The hoppings are now between different sites as well as different Floquet sectors. Additionally, in a given Floquet sector $m$, all the diagonal terms are shifted by $m\Omega$.

To calculate $t_{i,i+\boldsymbol{\delta}_{i}}^{\left(m\right)}$ for arbitrary polarization, we use that $\boldsymbol{E}=-\frac{\partial\boldsymbol{A}}{\partial t}$.  We can then simplify 
\begin{align}
\boldsymbol{\delta}_{l}\cdot\boldsymbol{A}\left(\theta\right) & =A_{l}\sin\left(\theta+\beta_{l}\right),\label{eq:dot-product-final}
\end{align}
where we define the direction dependent amplitude, $A_l$ and phase, $\beta_l$.
We have,
\begin{equation}
A_{l}=A_{0}\sqrt{I/I_{0}}\sqrt{1+\cos2\chi\cos\left[2\left(\psi-\phi_{l}\right)\right]}.\label{eq:A_l}
\end{equation}
where we introduce the average fluence $A_{0}=\frac{1}{\Omega}\sqrt{I_{0}/2}$.
Notice that $A_{l}$ is symmetric with respect to $\chi=0$, the case of linear
polarization. 
The phase $\beta_{l}$ is defined by
\begin{align}
\cos\beta_{l} & =\frac{\sqrt{2}\sin\chi\sin\left(\psi-\phi_{l}\right)}{\sqrt{1+\cos2\chi\cos\left[2\left(\psi-\phi_{l}\right)\right]}},\label{eq:cos-bl}\\
\sin\beta_{l} &=-\frac{\sqrt{2}\cos\chi\cos\left(\psi-\phi_{l}\right)}{\sqrt{1+\cos2\chi\cos\left[2\left(\psi-\phi_{l}\right)\right]}}.\label{eq:sin-bl}
\end{align}
Note that $\beta_l \rightarrow \pi - \beta_l$ as $\chi \rightarrow - \chi$. 
It is useful to examine the simpler cases of circular and linear polarization.
Circular polarization ($\chi = \pm \pi/4$) gives a direction-independent $A_{l}=A_{0}$, with
\begin{align}
 \beta_{l}=-\frac{\pi}{2}\pm\left(\psi-\phi_{l}\right)\,\,\,\,\,\,\,\,\,\text{(LCP/RCP)}\label{eq:beta-l-LCP-RCP}
\end{align}
For linear polarization, $\chi=0$, and
\begin{equation}
A_{l}=\sqrt{2}A_{0}\cos\left(\psi-\phi_{l}\right),\quad\beta_{l}=\pi/2.\,\,\text{(LP)}\label{eq:LP_simplified}
\end{equation}
We can now calculate $t_{i,i+\boldsymbol{\delta}_{i}}^{\left(m\right)}$ analytically for arbitrary $A_l$ and $\beta_l$, finding

\begin{align}
t_{i,i+\boldsymbol{\delta}_{i}}^{\left(m\right)} & =t_{1}\frac{1}{2\pi}\int_{0}^{2\pi}d\tilde{\theta}e^{-im\left(\tilde{\theta}-\beta_{l}\right)}\exp\left(-iA_{l}\sin\tilde{\theta}\right),\nonumber \\
 & =t_{1}e^{im(\beta_{l}+\pi)}\mathcal{J}_{m}\left(A_{l}\right)c_{i}^{\dagger}c_{i+\boldsymbol{\delta}_{i}}.\label{eq:tm}
\end{align}
We have used the change of variables, $\tilde{\theta}=\theta+\beta_{l}$, and the Bessel function representation $\mathcal{J}_{n}\left(x\right)=\frac{1}{2\pi}\int_{-\pi}^{\pi}e^{-i\left(n\theta+x\sin\theta\right)}d\theta.$  The hoppings have acquired both a directionally dependent amplitude, $t_1 \mathcal{J}_{m}\left(A_{l}\right)$ and complex phase $e^{-im\beta_{l}}$.  These hoppings satisfy,
\begin{equation}
t_{i+\boldsymbol{\delta}_{l},i}^{m-n}=\left(t_{i,i+\boldsymbol{\delta}_{l}}^{n-m}\right)^{*},
\end{equation}
which reduces to the expected  $t_{i+\boldsymbol{\delta}_{l},i}=t_{i,i+\boldsymbol{\delta}_{l}}^{*}$
if $m=n$.

The hopping can now be explicitly evaluated for linear
and circular polarization. For linear polarization, 

\begin{align}
t_{i+\boldsymbol{\delta}_{l},i}^{\left(m\right)} & =t_{1}e^{im\pi}\mathcal{J}_{m}\left(\sqrt{2}A_{0}\sqrt{I/I_{0}}\cos\left(\psi-\phi_{l}\right)\right)\,\,\text{(LP)}.\label{eq:LP-hops}
\end{align}
Here, we see that the overall phase is just $\pm 1$ and independent of $\phi_l$.  As all hoppings are real, there are no chiral fields generated.  The amplitude depends on the orientation of the nearest-neighbor link, $\phi_l$, which implies that the nearest-neighbor hoppings are now anisotropic. 

For circular polarization, the hopping is
\begin{equation}
t_{i+\boldsymbol{\delta}_{l},i}^{\left(m\right)}=t_{1}e^{\pm im\left(\psi-\phi_{l}\right)}\mathcal{J}_{m}\left(A_{0}\sqrt{I/I_{0}}\right).\,\text{\,\,\,(CP)}.\label{eq:hop-CP}
\end{equation}
 As expected, the hopping
amplitude $t_{1}\left|\mathcal{J}_{m}\left(A_{0}\sqrt{I/I_{0}}\right)\right|$
is independent of direction, as the polarization does not break lattice symmetries. However, the hopping is generically complex, with $\beta_{l}$ depending on $\mp \phi_l$ for RCP/LCP showing how the light helicity is
transferred to the Floquet hoppings. As the hoppings are complex,
the effective spin Hamiltonian generically breaks
time-reversal symmetry. Notice, however, that the phases depend on
the Floquet sector. As we show later, for the square lattice, selection rules for the allowed values of $m$ cause the time-reversal breaking terms vanish. 

\section{Calculating exchange couplings~\label{sec:General_results}}

Now that we have the modified hoppings, $t^{(m)}_{i+\boldsymbol{\delta}_{l},i}$, we can calculate the Floquet engineered exchange couplings.  In this section, we introduce the general calculation via Brillouin-Wigner perturbation theory and discuss the choice of materials, frequencies, and fluences to avoid heating issues while maximizing the tunability of the exchange couplings. Note that the formal derivation of the perturbation structure is quite technical and is thus left to Appendix~\ref{sec:Formal-structure-pert}. The second-order calculation for the interacting case has been extensively addressed before, both using Brillouin-Wigner and Schrieffer-Wolff approaches~\cite{PolkovnikovPRL2016,ClaassenNatComm2017,Refael_PRB_2019,BalentsPRL2018,BalentsPRB2018,Losada_PRB_2019}. Fourth-order expressions for circularly polarized light on the kagom\'e lattice were derived in Ref.~{\onlinecite{ClaassenNatComm2017}}.

\subsection{Perturbation theory}

In this subsection, we calculate the effective spin Hamiltonians emerging from the half-filled Floquet-Hubbard Hamiltonian for $t_1 \ll U \sim \Omega$.  Here, we assume that the frequency, $\Omega$ is comparable to the interaction, $U$, as this case allows resonances for $U \sim m \Omega$ that maximize the enhancement of the exchange couplings. 

In all calculations in this section, we keep the
polarization arbitrary, with any polarization averages 
performed later.

We can again decompose Eq.~(\ref{eq:Hamiltonian-honeycomb-Floquet}), into $\mathcal{H}=\mathcal{H}_{0}+\mathcal{V}$,
where the kinetic term $\mathcal{V}$ is the perturbation to $\mathcal{H}_{0}$
in the limit $t_{1}\ll U \sim \Omega$,

\begin{align}
\mathcal{H}_{0} & =\mathcal{H}_{\text{int}}+\sum_{m}m\,\Omega\left|m\right\rangle \left\langle m\right|\cr
\mathcal{V} & =\sum_{i,\boldsymbol{\delta}_{l}}\sum_{m,n}t_{i+\boldsymbol{\delta}_{l},i}^{\left(m-n\right)}c_{i,\sigma}^{\dagger}c_{i+\boldsymbol{\delta}_{l},\sigma}\left|n\right\rangle \left\langle m\right|.\label{eq:V-Floquet}
\end{align}
In order to find the effective spin models, we must systematically calculate higher order corrections in degenerate perturbation theory, for which we use the Brillouin-Wigner approach~\cite{lindgren1974BW,RaoPRB2016}.  For illustration purposes, we show some of the steps for the second order perturbation theory, 
leaving the details to Appendix~\ref{sec:Formal-structure-pert}.
The generic second-order correction reads

\begin{align}
\mathcal{\mathcal{H}}^{\left(2\right)} & =-\sum_{m}\left(P\mathcal{V}_{m}Q\right)\frac{1}{\left(U+m\Omega\right)}\left(Q\mathcal{V}_{-m}P\right),\label{eq:second-order-H}
\end{align}
Here, $P$ and $Q = 1-P$ are projectors into the ground and excited state manifolds, acting on the time-independent Hubbard model states, while $\mathcal{V}_m$ moves an electron from the ground state to an excited state with energy difference $U+m\Omega$.  Generically, $Q=\sum_k Q_{kU}$ encompasses all the excited states with energy $kU$, due to $k$ doubly occupied sites, but only a single doubly occupied intermediate state is involved to second order, $Q_U$.

\begin{figure}
\includegraphics[width=0.87\columnwidth]{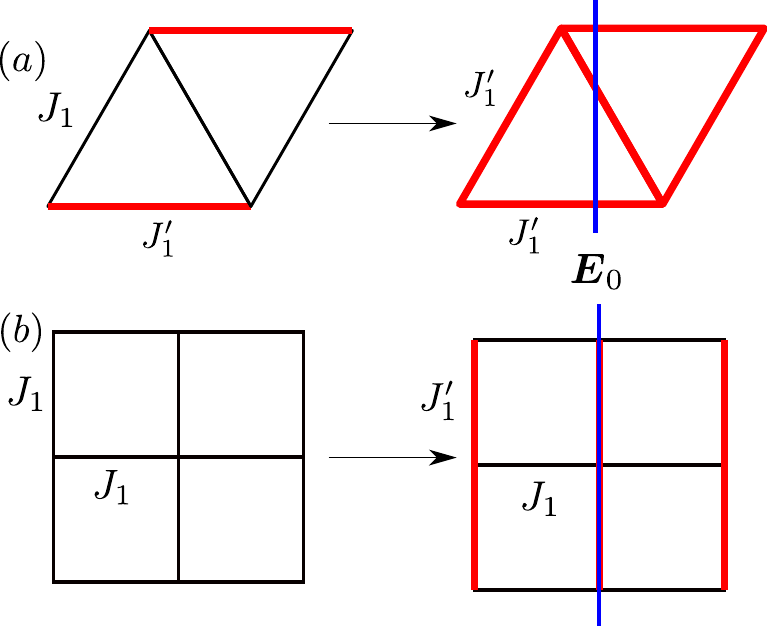}\caption{The original anisotropy of the equilibrium coupling constants, $J_1$ may be removed or enhanced by Floquet engineering with linearly polarized light.  Here, the vertical line indicates the electric field orientation.  (a) If the lattice is originally anisotropic, this anisotropy may be removed by applying linearly polarized light with the polarization vector perpendicular to the bond with $J_1' \neq J_1$, and tuning the fluence such that the modified $J_1$ given by Eq. (\ref{eq:J1-order-2}) is equal to the original $J_1'$. As the polarization is perpendicular to the $J_1'$ link, it is unaffected, and the nonequilibrium lattice will have \emph{isotropic} nearest-neighbor exchange couplings.  (b) If the lattice is originally isotropic and two-dimensional, it may be tuned towards one-dimensionality again by applying linearly polarized light to selectively enhance the couplings parallel to the polarization. \label{fig:anitropy-fixing}}
\end{figure}

We now must evaluate the operator product, $P\mathcal{V}_m Q \mathcal{V}_{-m} P$.  Similar products also appear in fourth-order calculations, so we evaluate the more general term here, $P\mathcal{V}_{-m_{3}}Q\mathcal{V}_{m_{3}-m_{2}}P$. Note that for $m_3 = -m$, $m_2 = 0$, the second order product is recovered. To evaluate this term explicitly, we consider only terms in $\mathcal{V}$ that move electrons from site $i$ to $i+\boldsymbol{\delta}_l$, and rewrite the projectors in terms of the spins, as usual~\cite{MacDonaldPRB1988,Tremblay2004PRB},
\begin{align}
P\mathcal{V}_{-m_{3}}^{\left(i,i+\boldsymbol{\delta}_{l}\right)}Q&\mathcal{V}_{m_{3}-m_{2}}^{\left(i+\boldsymbol{\delta}_{l},i\right)}P  =-\left(t_{i+\boldsymbol{\delta}_{l},i}^{\left(-m_{3}\right)}t_{i,i+\boldsymbol{\delta}_{l}}^{\left(m_{3}-m_{2}\right)}\right.\nonumber \\
&\left.+t_{i+\boldsymbol{\delta}_{l},i}^{\left(-m_{3}\right)}t_{i,i+\boldsymbol{\delta}_{l}}^{\left(m_{3}-m_{2}\right)}\right) \times\left(2\boldsymbol{S}_{i}\cdot\boldsymbol{S}_{j}-\frac{1}{2}\right).\label{eq:second-order-string}
\end{align}
This equation reduces to the well-known $J_{1}=\frac{4t_{1}^{2}}{U}$
when all the Floquet indices $m$ are taken to be zero. By using the
expression (\ref{eq:tm}) for the hoppings, the phases cancel and the exchange coupling becomes,

\begin{equation}
J_{1}^{\left(\boldsymbol{\delta}_{l}\right)}=4t_{1}^{2}\sum_{m}\frac{\left|\mathcal{J}_{m}\left(A_{l}\right)\right|^{2}}{\left(U+m\Omega\right)}.\label{eq:J1-order-2}
\end{equation}
This result is similar to Ref.~\onlinecite{ClaassenNatComm2017}
on the kagome lattice for circular polarization, as the lattice geometry and polarization
only enter $J_1^{\left(\boldsymbol{\delta}_{l}\right)}$ through $A_{l}$.  The further neighbor couplings, chiral fields and ring exchange terms are all more lattice/polarization dependent.  The generic exchange couplings are isotropic, as the polarization direction breaks the lattice symmetry, and can be manipulated to \emph{remove} equilibrium  anisotropy, as shown in Fig.~\ref{fig:anitropy-fixing}.

\begin{figure}[b!]

\includegraphics[width=1\columnwidth]{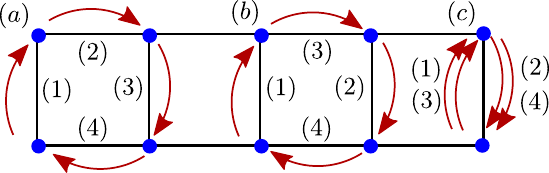}

\caption{Schematic of fourth-order processes on the square lattice. (1)-(4) represent the
four electron hops. (a) and (b) come from the
first and second terms of (\ref{eq:H4a}). (b) leads to
two doubly occupied sites, with a $2U$ denominator. (c)
represents either (\ref{eq:H4b}) or (\ref{eq:H4c}), depending
on the Floquet sectors involved. \label{fig:Schematic-perturbation-terms}}

\end{figure}

The formal derivation of the fourth-order terms is shown in  Appendix~\ref{sec:Formal-structure-pert}.  In general, there are four contributions to the perturbative Hamiltonian, as shown in Fig.~\ref{fig:Schematic-perturbation-terms} for the square lattice.  The fourth-order terms involve four
$\mathcal{V}$ operators, which implies that an electron can, at most,
hop to its third neighbor before coming back to its original site.
The spin of this electron can change during the process, and that
is the origin of the effective spin Hamiltonian. $\mathcal{\mathcal{H}}_{a,b,c}^{\left(4\right)}$
 indicate the three distinct terms in the fourth-order perturbation
theory {[}see Eq.~(\ref{eq:order-4-generic}) of Appendix~\ref{sec:Formal-structure-pert}{]}.  For simplicity, we package the $m$'s together as, $\textbf{m}\equiv (m_{1},m_{2},m_{3})$, and find the terms,

\begin{align}
\mathcal{\mathcal{H}}_{a}^{\left(4\right)} & =\!  -\sum_{\textbf{m}}\frac{P\mathcal{V}_{-m_{3}}Q_{U}\mathcal{V}_{m_{3}-m_{2}}Q_{U}\mathcal{V}_{m_{2}-m_{1}}Q_{U}\mathcal{V}_{m_{1}}P}{\left(U+m_{3}\Omega\right)\left(U+m_{2}\Omega\right)\left(U+m_{1}\Omega\right)}\nonumber \\
 & -\sum_{\textbf{m}}\frac{P\mathcal{V}_{-m_{3}}Q_{U}\mathcal{V}_{m_{3}-m_{2}}Q_{2U}\mathcal{V}_{m_{2}-m_{1}}Q_{U}\mathcal{V}_{m_{1}}P}{\left(U+m_{3}\Omega\right)\left(2U+m_{2}\Omega\right)\left(U+m_{1}\Omega\right)}\label{eq:H4a}\\
\mathcal{\mathcal{H}}_{b}^{\left(4\right)}&=- \!\!\!\!\! \sum_{\textbf{m} (m_{2}\ne0)}\!\!\!\!\frac{P\mathcal{V}_{-m_{3}}Q_{U}\mathcal{V}_{m_{3}-m_{2}}P\mathcal{V}_{m_{2}-m_{1}}Q_{U}\mathcal{V}_{m_{1}}P}{\left(U+m_{3}\Omega\right)\left(m_{2}\Omega\right)\left(U+m_{1}\Omega\right)},\label{eq:H4b}\\
\mathcal{\mathcal{H}}_{c}^{\left(4\right)}&= \sum_{m_{1},m_{2}}\frac{P\mathcal{V}_{-m_{2}}Q_{U}\mathcal{V}_{m_{2}}P\mathcal{V}_{-m_{1}}Q_{U}\mathcal{V}_{m_{1}}P}{\left(U+m_{2}\Omega\right)^{2}\left(U+m_{1}\Omega\right)}.\label{eq:H4c}
\end{align}
These comprise all corrections from fourth-order perturbation theory.  
 Notice that the string of operators in the
numerator of Eqs.~(\ref{eq:H4b}) and (\ref{eq:H4c}) are identical,
only differing by $m$'s. As each of these involves multiple sites, the challenge is to evaluate the products of projectors, which depend strongly upon the lattice geometry.  There are, however, four generic functions that recur in the specific lattice calculations.  At this point, it is convenient to define the dimensionless ratios, $\tilde{t} = t_1/U$, $\tilde{\Omega} = \Omega/U$.

\begin{widetext}

\begin{align}
\mathcal{A}_{ijkl}\left(\boldsymbol{m}\right) & =\left(-1\right)^{m_{2}}\tilde{t}^{3}\frac{\mathcal{J}_{-m_{3}}\left(A_{l_{i}}\right)\mathcal{J}_{m_{3}-m_{2}}\left(A_{l_{j}}\right)\mathcal{J}_{m_{2}-m_{1}}\left(A_{l_{k}}\right)\mathcal{J}_{m_{1}}\left(A_{l_{l}}\right)}{\left(1+m_{1}\tilde{\Omega}\right)\left(1+m_{2}\tilde{\Omega}\right)\left(1+m_{3}\tilde{\Omega}\right)},\label{eq:A_m}\\
\mathcal{L}_{ijkl}\left(\boldsymbol{m}\right) & =(-1)^{m_{1}+m_{3}}\tilde{t}^{3}\cos^{2}\left(m_{2}\frac{\pi}{2}\right)\frac{\mathcal{J}_{-m_{3}}\left(A_{l_{i}}\right)\mathcal{J}_{m_{3}-m_{2}}\left(A_{l_{j}}\right)\mathcal{J}_{m_{2}-m_{1}}\left(A_{l_{k}}\right)\mathcal{J}_{m_{1}}\left(A_{l_{l}}\right)}{\left(1+m_{1}\tilde{\Omega}\right)\left(2+m_{2}\tilde{\Omega}\right)\left(1+m_{3}\tilde{\Omega}\right)},\label{eq:L_m}\\
\mathcal{B}_{ij}\left(\boldsymbol{m}\right) & =\left(-1\right)^{m_{1}+m_{3}}\tilde{t}^{3}\cos^{2}\left(m_{2}\frac{\pi}{2}\right)\frac{\mathcal{J}_{-m_{3}}\left(A_{l_{i}}\right)\mathcal{J}_{m_{3}-m_{2}}\left(A_{l_{i}}\right)\mathcal{J}_{m_{2}-m_{1}}\left(A_{l_{j}}\right)\mathcal{J}_{m_{1}}\left(A_{l_{j}}\right)}{\left(1+m_{1}\tilde{\Omega}\right)\left(m_{2}\tilde{\Omega}\right)\left(1+m_{3}\tilde{\Omega}\right)},\,\,m_{2}\ne0,\label{eq:B_m}\\
\mathcal{G}_{ij}\left(\boldsymbol{m}\right) & =\tilde{t}^{3}\delta_{m_{2},0}\left[\mathcal{J}_{m_{1}}^{2}\left(A_{l_{i}}\right)\mathcal{J}_{m_{3}}^{2}\left(A_{l_{j}}\right)+\mathcal{J}_{m_{1}}^{2}\left(A_{l_{j}}\right)\mathcal{J}_{m_{3}}^{2}\left(A_{l_{i}}\right)\right]\frac{1}{\left(1+m_{1}\tilde{\Omega}\right)^{2}\left(1+m_{3}\tilde{\Omega}\right)}.\label{eq:G_m}
\end{align}

\end{widetext} 

\begin{figure}
\includegraphics[width=0.8\columnwidth]{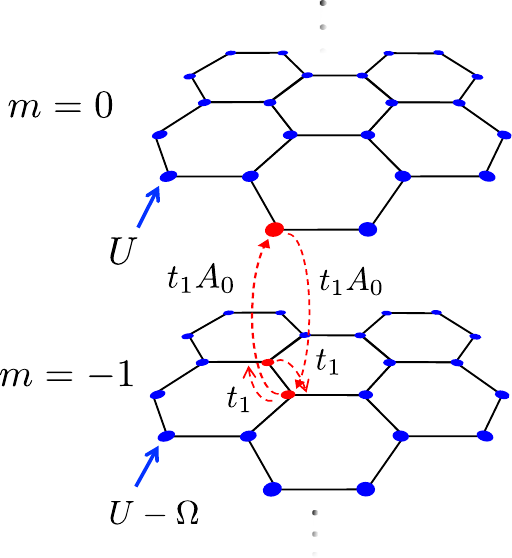}

\caption{Fourth order corrections to $J_1$ and $J_2$ on the honeycomb lattice involve processes like the one shown here, where the electrons hop between different sites and Floquet sectors. Here, for simplicity, we expand the inter-sector hoppings $t_{1}J_{m}\left(A_{0}\right)\sim t_{1}A_{0}^{m}$, which is true for $A_0 \ll 1$.  In the excited, $m=-1$ sector, the Coulomb interaction is lowered to $U-\Omega$. Here we show a process in which the electron hops around the $m=-1$ sector before returning to the original site. The full calculation involves electrons hopping to arbitrary Floquet sectors. \label{fig:pert_processes}}
\end{figure}

These functions do not include any projectors, as these are converted to expressions involving the spins, like $\left(2\boldsymbol{S}_{i}\cdot\boldsymbol{S}_{j}-\frac{1}{2}\right)$; these are simply the relevant coefficients, incorporating the renormalized hoppings as well as the energy denominators. The indices
$i,j,k,l$ label the hopping directions along the lattice. $\mathcal{A}_{ijkl}$ and $\mathcal{L}_{ijkl}$ come from
from the two terms of Eq.~(\ref{eq:H4a}), while $\mathcal{B}_{ij}$ comes from (\ref{eq:H4b}) and $\mathcal{G}_{ij}$ from (\ref{eq:H4c}).  As any arbitrary polarization breaks lattice symmetry, these functions really do depend on the sites involved and lead to anisotropic exchange couplings that depend on $i,j,k,l$.  For circular polarization, and after polarization averaging for the different kinds of unpolarized light, these will wash out.

An important check is that the exchange couplings so derived match the time-independent (bare) case for $A_0 \rightarrow 0$, which forces $\boldsymbol{m}=\boldsymbol{0}$, at which point the limit $\Omega\rightarrow0$ can be safely taken. The expressions
for $J_{1}$, $J_{2}$, $J_{3}$ and $J_{\square}$ in the absence
of Floquet fields are shown for the three lattices studied in this
paper in Table~\ref{tab:exchange-time-independent}. 

By looking at the fourth-order expressions one might infer that, in higher orders in perturbation theory, the generic denominators will be of the form $n U + m \Omega$. These would correspond to $m$ photons exciting $n$ pairs of holons/doublons.  Interference of different paths, however, restricts the resonances only to $\Omega/U=1/\tilde{m}$, with $\tilde{m}$ integer. This is not obvious from the Brillouin-Wigner approach used here, but it is evident in the Schrieffer-Wolff formulation~\cite{SchriefferWolfforiginal,PolkovnikovPRL2016}, which should yield exactly the same results as the Brillouin-Wigner approach for any given order in perturbation theory.

\begin{center}
\begin{table}
\begin{centering}
\begin{tabular}{|c|c|c|c|}
\hline 
 & Honeycomb & Square & Triangular\tabularnewline
\hline 
\hline 
$J_{1}$ & $4\frac{t_{1}^{2}}{U}-16\frac{t_{1}^{4}}{U^{3}}$ & $4\frac{t_{1}^{2}}{U}-24\frac{t_{1}^{4}}{U^{3}}$ & $4\frac{t_{1}^{2}}{U}-28\frac{t_{1}^{4}}{U^{3}}$\tabularnewline
\hline 
$J_{2}$ & $4\frac{t_{1}^{4}}{U^{3}}$ & $4\frac{t_{1}^{4}}{U^{3}}$ & $4\frac{t_{1}^{4}}{U^{3}}$\tabularnewline
\hline 
$J_{3}$ & - & $4\frac{t_{1}^{4}}{U^{3}}$ & $4\frac{t_{1}^{4}}{U^{3}}$\tabularnewline
\hline 
$J_{\square}$ & -- & $80\frac{t_{1}^{4}}{U^{3}}$ & $80\frac{t_{1}^{4}}{U^{3}}$\tabularnewline
\hline 
\end{tabular}
\par\end{centering}
\caption{Time-independent exchange couplings up to fourth order for the lattices addressed.
$J_{1},J_{2}$ and $J_{3}$ correspond to nearest, second and third neighbor
couplings on each lattice. $J_{\square}$ is ring-exchange.  There are no chiral fields here, as time-reversal is preserved. Note that the factor of 80
in $J_{\square}$ may look large, but comes from the
use of spin matrices, which satisfy
$S_i^{2}=1/4$. The plaquette term is quartic in spin operators,
while the other terms are quadratic.
\label{tab:exchange-time-independent}}
\end{table}
\par\end{center}

\subsection{Resonances, heating and connection to experiments~\label{sec:Connection-to-experiments}}

In order to successfully Floquet engineer a material into a new state of matter, it is essential that:
\begin{itemize}
    \item There exists a transient, pre-thermalized Floquet regime, in which not only are the exchange coupling modified, but the system relaxes into the state favored by these new couplings, without heating the system to temperatures high enough to wash out the physics of interest. In our case, as the materials are insulating, heating can be substantially avoided by avoiding exciting electrons between the upper and lower Hubbard band.
    \item The new state must then be characterized during the short time-scales of the Floquet pulse, which rules out most conventional magnetic measurements.  Optical techniques are ideally compatible with the pump-probe nature of Floquet experiments. Discontinuities associated with phase transitions should be observable in optical measurements, and magnetic excitations can be followed~\cite{Banerjee2018}.  Spin liquids have neutral low energy spinon excitations that couple only weakly to the external gauge field, but gapless spin liquids have been predicted to have power law behaviors in optical conductivity~\cite{PotterPRB2013,pilon13,pustogow18}, and spin liquids, in general, may have signatures in the magneto-optical Faraday or Kerr effects~\cite{colbert14}.
\end{itemize}

\begin{figure}
\includegraphics[width=0.75\columnwidth]{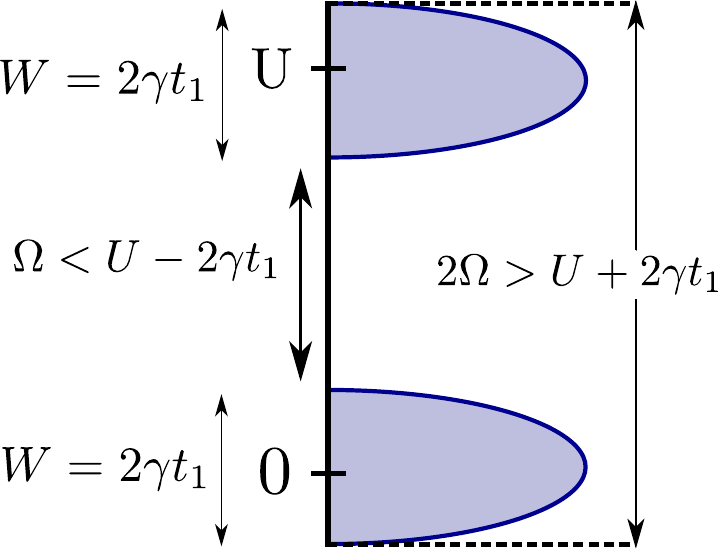}\caption{Cartoon of the two lowest Hubbard bands, and the constraints
they impose upon the frequency. We focus on $\Omega$ in the range $1/2\,U$ to $U$, where $\Omega$ must simultaneously satisfy two conditions: $\Omega<U-2\gamma t_1$ to avoid the excitation of pairs of doublons and holons between the top
of the lower Hubbard band and the bottom of the top Hubbard band; and $2\Omega>U+2\gamma t_1$, which prevents a pair of photons exciting doublon-holon pairs  from the bottom of the lower band to the top of the upper band. These requirements restrict $\Omega$ significantly. For the triangular lattice, for instance, if $t_1/U>\frac{1}{12\sqrt{5}}\approx 0.0372$, there is
no $\Omega/U$ satisfying these requirements. \label{fig:Hubbard-bands}}
\end{figure}

\begin{figure}
\includegraphics[width=1\columnwidth]{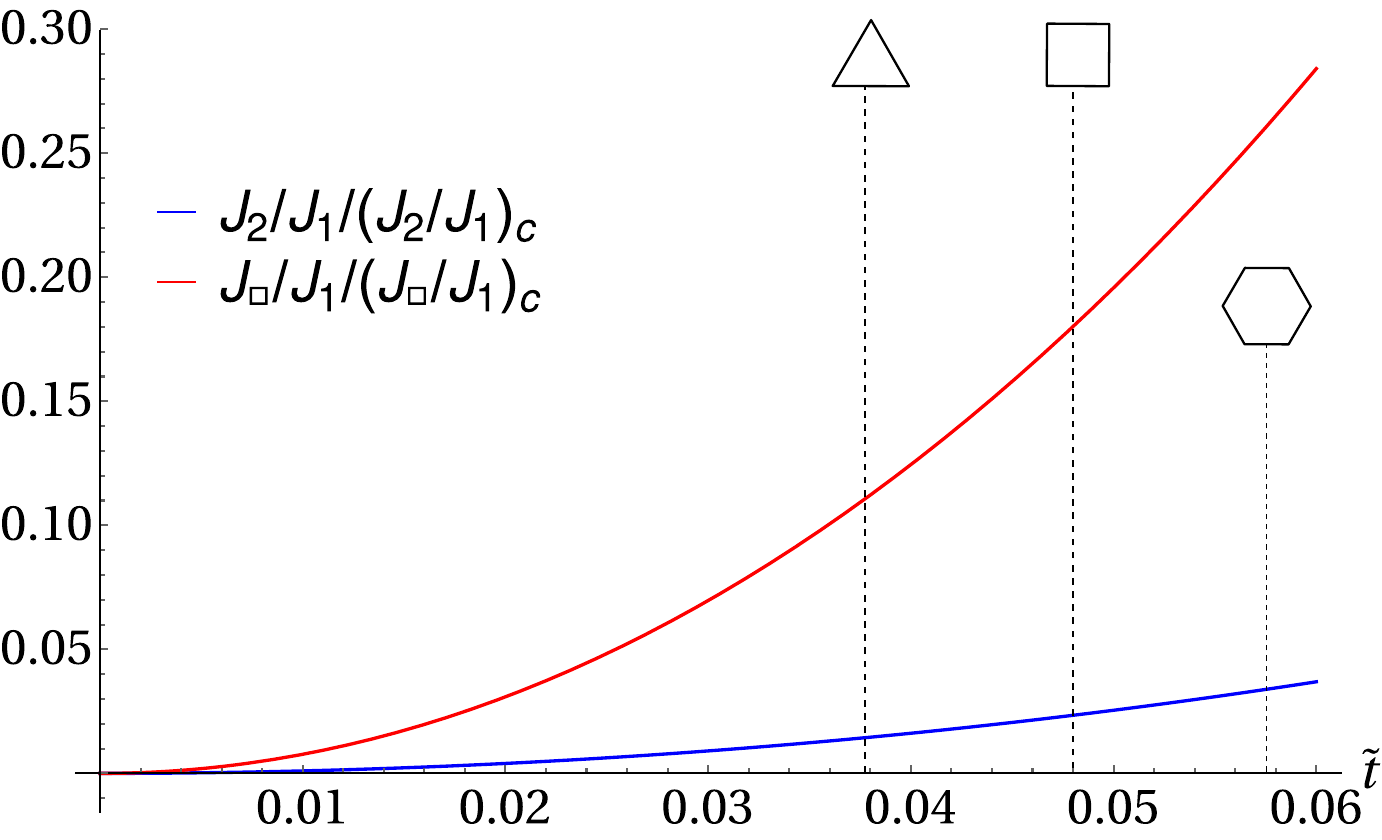}\caption{$J_{2}/J_{1}$ and $J_{\square}/J_{1}$ versus $\tilde{t}$ for the time-independent problem, based on Table~\ref{tab:exchange-time-independent}, and scaled by the critical values for the triangular lattice, $(J_2/J_1)_c = 0.1$ and $(J_\square/J_1)_c =0.2$. Note that the triangular lattice critical values are by far the lowest of the three lattices, and so it is clear that the time-independent problem is far from the critical points. The choice of $\tilde{t}$ in this range $0-\frac{1}{12\sqrt{z-1}}$ guarantees that the starting point is sufficiently deep inside the Mott insulating phase that heating can be avoided.  $J_{1}$ has different sub-leading fourth-order corrections depending on the lattice, but these contributions are not enough to change $J_{1}$ substantially in this range of $\tilde{t}$. $J_{3}$ is also present and assumes the same value as $J_{2}$. Both $J_{3}$ and $J_{\square}$ are absent on the honeycomb lattice to fourth order. 
\label{fig:Hubbard_starting_point}}
\end{figure}

Deep in the Mott insulating regime, the hopping $\tilde{t}$ will be negligible and the excited states of the material will simply be a set of discrete levels separated by $U$, representing $k$ doubly occupied sites.  Increasing $\tilde{t}$ allows electrons to hop, changing which sites are empty/doubly occupied (holons/doublons) and broadening the discrete levels by some finite bandwidth, $W = 2 \gamma t_1$, as shown roughly in Fig.~\ref{fig:Hubbard-bands}.  In order to avoid heating, it must not be possible to excite electrons between these discrete levels, with any number of photons.  We find the greatest enhancements when the frequency is in the region $U/2< \Omega < U$, in large part because the $m$-th Bessel functions are involved in processes involving $m$ photons (or $U/m$ resonances), which tend to decrease as $m$ increases. Here, a single photon must not be able to excite a  doublon-holon pair from the top of the lower Hubbard band to the bottom of the upper Hubbard band, $\Omega < U - W$. In addition, two photons must not be able to excite a doublon-holon pair from the bottom of the lower to the top of the upper band, $2 \Omega > U + W$.  These two requirements combined,
\begin{equation}
\Omega<U-W,\,\,\,2\Omega>U+W.\label{eq:heating-constraints}
\end{equation}
are sufficient to ensure that no number of photons can excite doublon-holon pairs across the gap, and thus avoid heating.  These requirements also enforce a maximum $\tilde{t} = 1/(6\gamma)$, beyond which there is no frequency between $U/2$ and $U$ that will not induce heating, which quickly heats the system to infinite temperatures.  Frequencies between other resonances, e.g., $U/3 < \Omega < U/2$ are even more restrictive.  It is always possible to find $\Omega > U +2 W$ that does not heat the system, however, the enhancements of the coupling constants here are typically quite small.  As such, we will restrict our $\tilde{t} < 1/(6\gamma)$, which implies
\begin{align}
\tilde{t}_{\text{max}} & =0.0589\,\,\left(\text{honeycomb}\right),\\
\tilde{t}_{\text{max}} & =0.0481\,\,\left(\text{square}\right),\\
\tilde{t}_{\text{max}} & =0.0372\,\,\left(\text{triangular}\right).
\end{align}
For the lattices studied in this work, we take the approximation for $\gamma$ from Ref.~\onlinecite{BalentsPRL2018}, $\gamma=2 \sqrt{z-1}$, with $z$ the coordination number of each lattice, $z=3$ for the honeycomb lattice, $z=4$ for the square lattice and $z=6$ for the triangular lattice.  In  Fig.~\ref{fig:Hubbard_starting_point}, we show the maximum initial (time-independent) values of $J_2/J_1$ and $J_\square/J_1$ possible for each of the three lattices.
Note that this requirement likely rules out the organic triangular lattice spin liquid candidates, which are generally close to the metal-insulator transition~\cite{shimizu03,Yamashita2008,Yamashita2008B}.  The denominators in the magnetic exchange couplings will be smallest when $\Omega = U/2 + \gamma t_1$, with the maximum enhancement occurring for $\tilde{t} = \tilde{t}_{\text{max}}$ and $\tilde{\Omega} = 2/3$.

While the condition above is necessary to avoid pairs of doublons and holons, other, less destructive mechanisms of heating from phonons or other collective modes will be present to some degree.  These mechanisms, expected to be relevant for real materials, are beyond the simple Hubbard model considered in this work. In general, a possible way of avoiding heating is by connecting the system to a heat bath, but the mechanisms of heating transfer must be studied case-by-case.

Next, we consider the experimental feasibility of reaching the appropriate frequencies and fluences.  The frequency should be $\sim 2/3 U$, on the order of the Mott gap, which is typically in the range $1-10$eV.  The dimensionless fluence required to maximize the enhancements is typically of order one, as the renormalized hoppings depend on Bessel functions that oscillate and decay for larger arguments.  The dimensionless fluence, in terms of dimensionful quantities, reads

\begin{equation}
A_{0}=\frac{a_{0}eE}{\hbar\Omega},
\end{equation}
where $a_{0}$ is the lattice spacing, on the order of Angstroms.
The intensity, with full units, reads

\begin{align}
I & =c\epsilon_{0}\left(\frac{\Omega\hbar}{ea_{0}}\right)^{2}\left|A_{0}\right|^{2},\nonumber \\
 & \approx 2.6\times10^{17}\left(\frac{\Omega\hbar\left[eV\right]}{a_{0}\left[\text{\AA}\right]}\right)^{2}\left|A_{0}\right|^{2}W/m^{2}
\end{align}
with $\epsilon_{0}$ the vacuum permittivity. The field strength $eE$
available varies depending on the experiment, ranging from $\left(0.01-1\right)eV/\text{\AA}\,$~\cite{Huber2008OptLett,Wang2013Science},
leading to intensities of $I\approx10^{15}-10^{17}W/m^{2}$. This
gives an order of magnitude estimation for $A_{0}$ in the region
between $0.01$ and $1$.  Our optimal fluences are typically $A_0 
\sim 1-3$, which seems reasonable for current experimental set ups.  However, keep in mind that relatively long pulse times might be required to create unpolarized light, which reduces the available fluence.

\section{Results for specific lattices~\label{sec:Specific_lattices}}

We now are ready to examine how the exchange couplings can be manipulated on three common two-dimensional lattices, which we will approach in order of difficulty, or number of nearest-neighbors: honeycomb ($z=3$), square ($z=4$) and triangular ($z=6$).  The honeycomb lattice only has two new terms arising at fourth order: $J_2$ and a chiral field, $J_\chi$, while the chiral term vanishes on the square lattice, but third neighbor and ring exchange terms are added.  The triangular lattice has all four couplings, with two distinct, although proportional chiral fields.

\subsection{Honeycomb lattice~\label{sec:Honeycomb_lattice}}

The honeycomb lattice is simplest not only because $z = 3$, but also because there are no closed loops to fourth order, and thus no ring exchange terms.  Here, we explore how $J_1$, $J_2$ and $J_\chi$ couplings are generated.  $J_\chi$ is strictly zero for unpolarized light and maximized for fully circularly polarized light, while $J_2$ is induced for all types of light.  Here, we explore how $J_2/J_1$ and $J_\chi/J_1$ can be tuned. We restrict ourselves to polarizations that do not break lattice symmetries by considering circular polarization, type I light that samples the whole Poincar\'e sphere evenly, and all types of type II light ($\chi = \pm \chi_0$), including type II Glauber ($\chi = 0$).

\begin{figure}
\includegraphics[width=1\columnwidth]{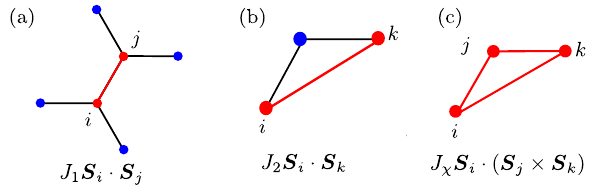}\caption{Representation of the sites involved in the exchange couplings to
fourth-order on the honeycomb lattice. The sites in the final expression are
represented in red, while the other sites are shown in blue.
(a) $J_{1}$ (b) $J_{2}$ (c) $J_{\chi}$. \label{fig:all-paths-honeycomb}}
\end{figure}

Calculating the exchange couplings up to fourth order involves hopping to a number of neighboring sites, as shown  in Fig.~\ref{fig:all-paths-honeycomb}.  The nearest-neighbor coupling, $J_1$ between sites $i,j$ involves four other neighboring sites, while $J_2$ involves only one intermediate site. 
As the calculation is tedious, in Appendix~\ref{sec:Floquet-Hubbard-pert} we derive the fourth order terms on the honeycomb lattice, and give the complete expressions in  Appendix~\ref{sec:Expressions}, up to fourth order. The calculations are straightforward, but involve a large number of paths that makes it more convenient to perform the calculations in algebraic software.  

\begin{figure}[htbp]
\includegraphics[width=1\columnwidth]{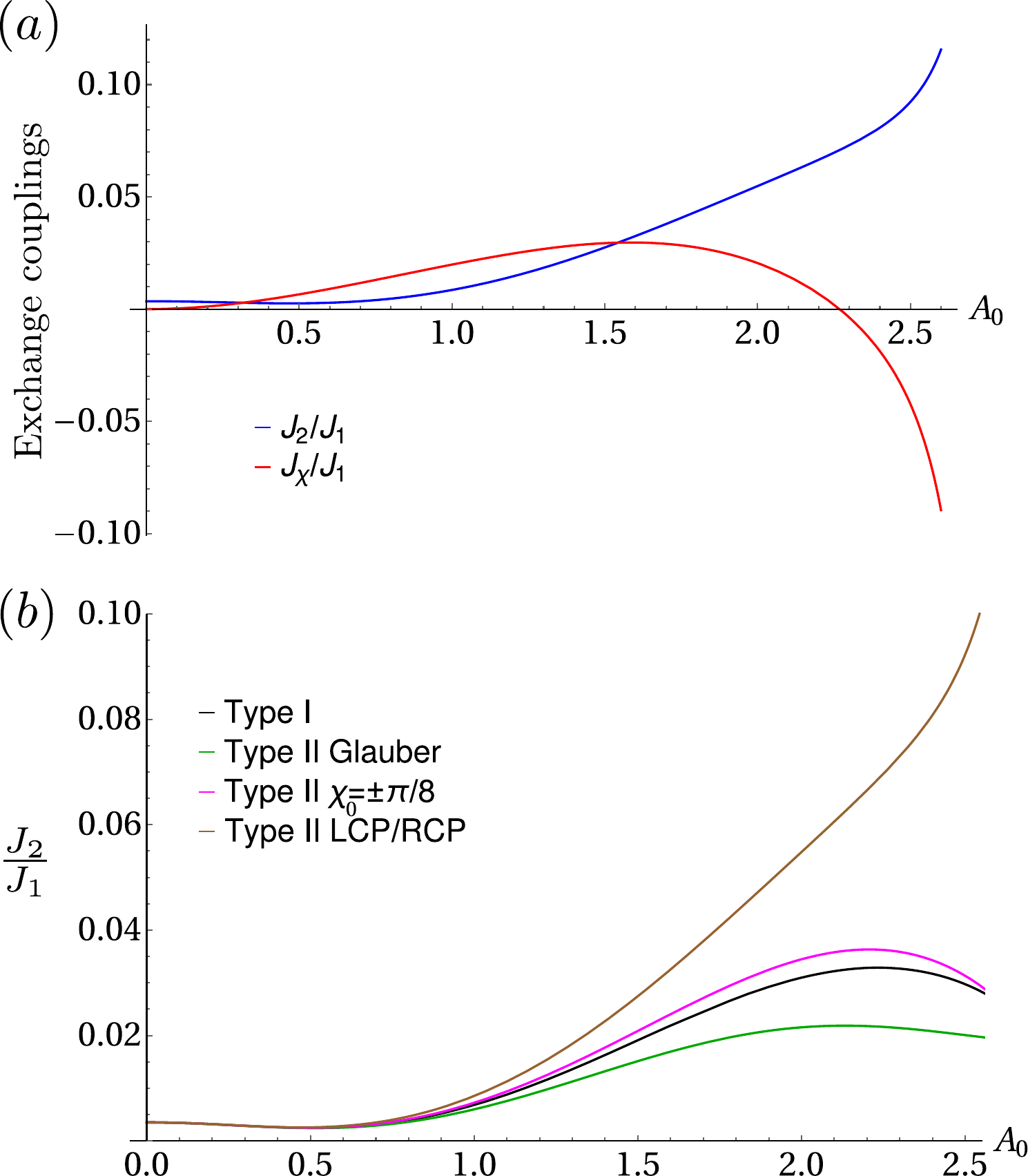}

\caption{$J_{2}/J_{1}$ and $J_{\chi}/J_{1}$ as function of $A_{0}$, for the maximum $\tilde{t}=0.059$ and $\tilde{\Omega}=2/3$, and for both (a) circularly polarized light and (b) Unpolarized light of different types.   These are cuts from Figs.~\ref{fig:honeycomb_lattice_plots}
and \ref{fig:honeycomb_lattice_plots-unpolarized}, respectively. The plots are stopped at $A_0 =2.5$, as this point is where the $J_1$ terms become very small and fourth-order perturbation theory is insufficient.  Note that the maximum $J_2/J_1$ is around $0.1$, for circular polarization, while the type II Glauber light is much less effective on the honeycomb lattice.  $J_\chi$ can be fine-tuned to zero, even for finite fluence, but the maximum $|J_\chi/J_1|$ is still $\sim 0.1$, too small to induce a chiral spin liquid.
 \label{fig:honeycomb_lattice_cut}}
\end{figure}

For arbitrary polarization, the couplings will generically depend on the two or three sites involved,  $J_{2}^{\left(i,k\right)}$ and $J_{\chi}^{\left(i,j,k\right)}$.
For concreteness, in what follows, we take the sites $i,j,k$ positioned
according to Fig. \ref{fig:all-paths-honeycomb}. The hoppings connecting
$i$ and $k$ are, therefore, along the directions $\boldsymbol{\delta}_{1}$
and $\boldsymbol{\delta}_{2}$, with other directions obtained similarly.  The general expressions for $J_2$ and $J_\chi$ are shown in Eqs.~(\ref{eq:J2-honey}) and (\ref{eq:Jchi-honey}), for arbitrary polarization.  The case of circular polarization reproduces the $J_2$ and $J_\chi^{(honeycomb)}$ results from  Ref.~\onlinecite{ClaassenNatComm2017}, as the geometry is identical for these two couplings.  $J_1$, however, is different.  The results simplify if the angle $\chi$ is set to zero and $\psi$ averaged over to give the type II Glauber results,
\begin{align}
\left\langle J_{2}^{\left(i,k\right)}\right\rangle  & =\sum_{\boldsymbol{m}}-4\left\langle \mathcal{A}_{2,1,2,1}\left(\boldsymbol{m}\right)\right\rangle +8\left\langle \mathcal{B}_{1,1,2,2}\left(\boldsymbol{m}\right)\right\rangle +\nonumber \\
 & +4\left\langle \mathcal{G}_{2,1}\left(\boldsymbol{m}\right)\right\rangle ,\label{eq:J2-honeycomb-LP}\\
J_{\chi}^{\left(i,j,k\right)} & =0.
\end{align}
Here, $J_{\chi}^{\left(i,j,k\right)}$ vanishes as a consequence of $\beta_{2}-\beta_{1}=0$ for linear polarization (see Eq.~(\ref{eq:LP_simplified})), or more straightforwardly because linear polarization preserves time-reversal symmetry. In Fig.~\ref{fig:honeycomb_lattice_cut}, we show how these terms vary as a function of fluence for different polarization protocols, normalized by $J_1$ in order to compare to theoretical values.

\begin{figure}[htbp]
\includegraphics[width=1\columnwidth]{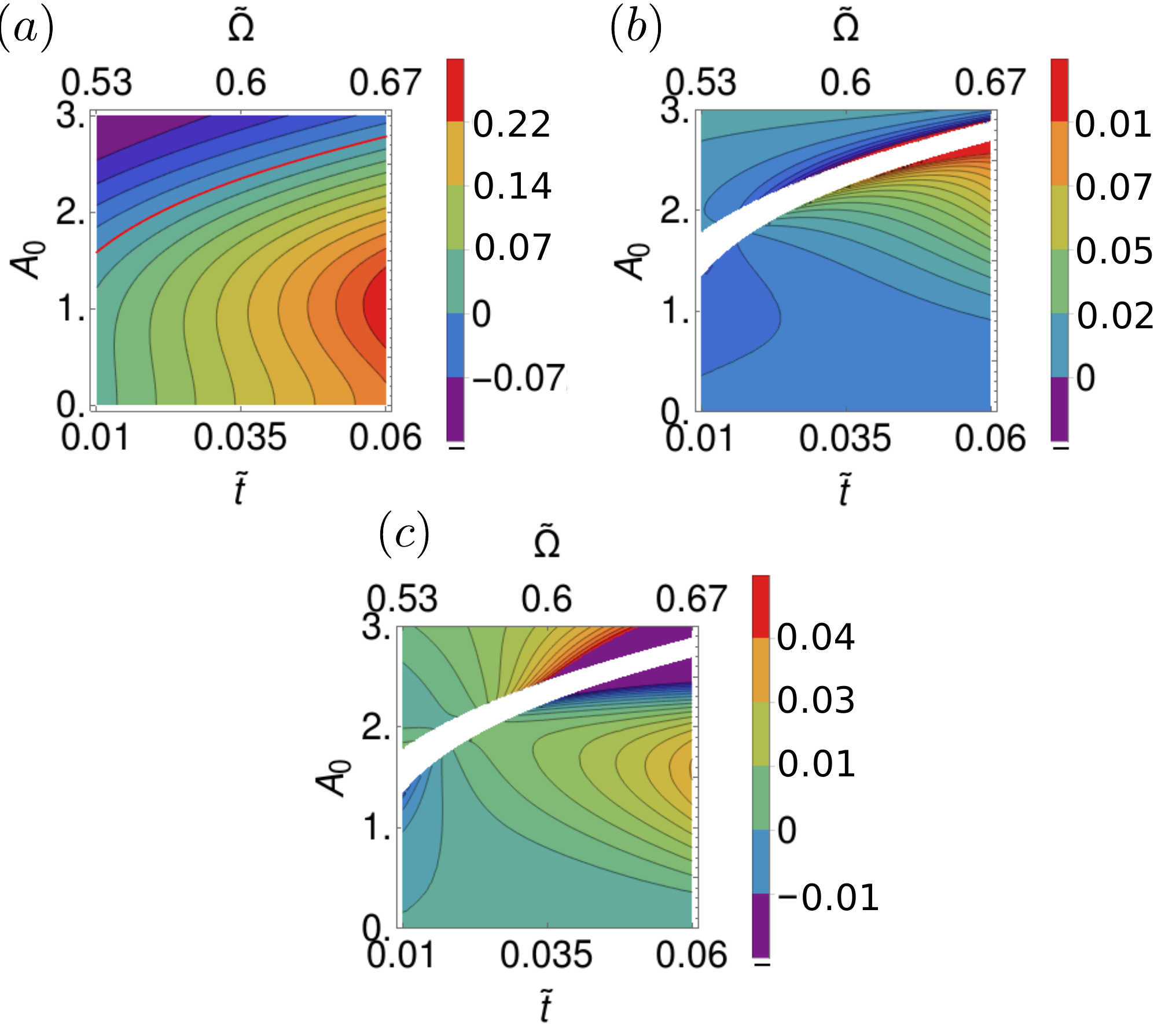}\caption{The couplings (a) $J_{1}$, (b) $J_{2}/J_{1}$ and (c) $J_{\chi}/J_{1}$
on a honeycomb lattice coupled to circularly polarized light. These are shown as
functions of $\tilde{t}$ and $A_{0}$, with the frequency, $\tilde{\Omega}=1/2+2\sqrt{2}\tilde{t}$, close to the  $\tilde{\Omega} = 1/2$ resonance.
In (a), the red line shows the curve $J_{1}=0$. Perturbation theory breaks down near this line, and in (b) and (c), we exclude $\left|J_{1}\right|<0.01$
(white regions). Both $J_{2}/J_{1}$ and $J_{\chi}/J_{1}$
may be significantly enhanced, but not enough to drive the system out of the N\'eel
phase, at least not without $J_1$ vanishing to destroy the N\'eel order in a more trivial way. \label{fig:honeycomb_lattice_plots}}
\end{figure}

The expression for the fourth-order contribution to $J_{1}$, which
we call $\delta J_{1}$ is unwieldy for arbitrary polarization, but we can give the relatively simpler expressions for circular and linear polarizations in Appendix~\ref{sec:Expressions}.
$J_1$ has both second and fourth order contributions, given by Eqs.~(\ref{eq:J1-order-2}) and (\ref{eq:J1-honey}). The
fourth-order corrections are almost always significantly smaller than the second-order contributions, given that we take $\tilde{t}$ to be small.  However, there is a region where $J_1$, as calculated in fourth order perturbation theory, becomes vanishingly small and passes through zero.  This behavior is primarily due to the second order contributions decreasing substantially; for sufficiently small second and fourth order terms, when these terms are comparable, the sixth order contributions must be considered, and so we omit the region where $J_1$ becomes this small from our other plots.  See Appendix~\ref{sec:J1-order2-order4} for more details about the validity of the perturbation expansion.

\begin{figure}[htbp]
\includegraphics[width=1\columnwidth]{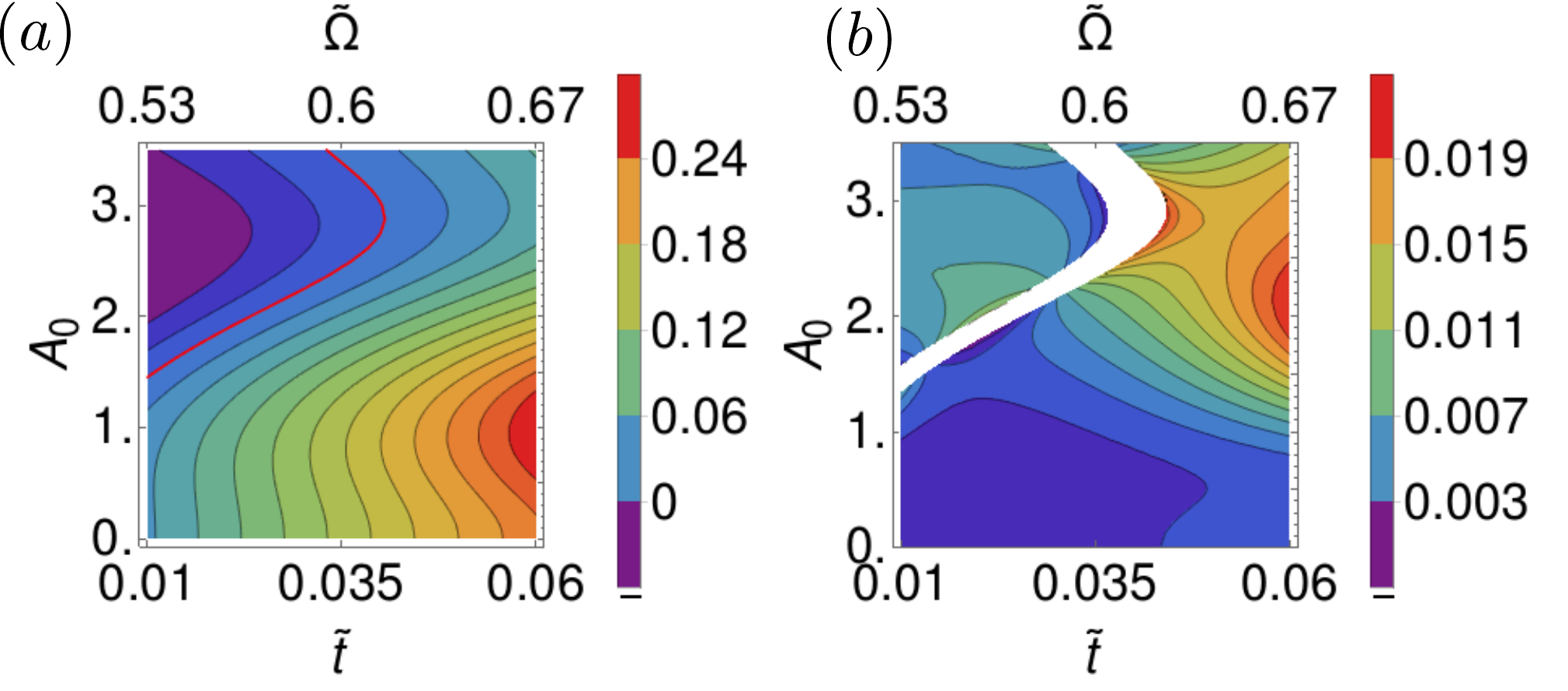}\caption{Exchange couplings on the honeycomb lattice, (a) $J_{1}$ (b) $J_{2}/J_{1}$, for type II Glauber unpolarized light.  These are shown as
functions of $\tilde{t}$ and $A_{0}$, with the frequency, $\tilde{\Omega}=1/2+2\sqrt{2}\tilde{t}$,  close to the  $\tilde{\Omega} = 1/2$ resonance.
In (a), the red line shows the curve $J_{1}=0$. Perturbation theory breaks down near this line, and in (b), we exclude $\left|J_{1}\right|<0.01$
(white regions). 
\label{fig:honeycomb_lattice_plots-unpolarized}}
\end{figure}

As $\tilde{t}$ is a materials property that cannot easily be tuned, we present our results as contour plots in Figs.~\ref{fig:honeycomb_lattice_plots} and \ref{fig:honeycomb_lattice_plots-unpolarized} as a function of $\tilde{t}$ and the dimensionless fluence $A_0$.  Here, we have chosen the frequency $\tilde{\Omega}$ that pushes the material as close to the $\tilde{\Omega} = 1/2$ resonance as possible without problematic heating; this value is $\tilde{\Omega} = 1/2+2\sqrt{2}\tilde{t}$. The line where $J_1$ vanishes (up to fourth order) is indicated in red, and a region around that line is omitted from the plots of $J_2/J_1$ and $J_\chi/J_1$.

The honeycomb lattice is bipartite, and so it takes a fairly substantial $J_2/J_1 \sim 0.2$ to induce a transition from the N\'eel state into either a spin liquid~\cite{ClarkPRL2011,gong_PRB_2019} or, more likely a plaquette valence bond solid phase via a deconfined critical point~\cite{Albuquerque2013PRB,GaneshPRL2013,Senthil2004PRB}. There are a few potential materials realizing the $S=1/2$ honeycomb lattice~\cite{MasatoshiPSJ2006,FreimuthJMMM2005,Cava2012IC,ChenPRB2012}, but $J_2/J_1$ is typically quite small, around $0.02$~\cite{ChenPRB2012}.  Here we will show that Floquet engineering the single-band Hubbard model can theoretically give a maximum $J_2/J_1 \sim 0.1$, about five times larger than those found in materials, and six times larger than the initial, time-independent values of our model.  Unfortunately, this value is only 50\% of the critical $J_2/J_1$, but it is possible that a material with preexisting $J_2/J_1 \gtrsim 0.1$ could be driven through the critical value via Floquet engineering.  The honeycomb lattice also hosts a potential chiral spin liquid, requiring up to third neighbors, and $J_\chi/J_1 \sim 0.25$~\cite{ParamekantiPRL2016}.  The maximum $J_\chi/J_1 \sim 0.04$ possible here is too small to induce a transition, and so the chiral spin liquid is out of reach, as there are no other sources of $J_\chi$. 

\subsection{Square lattice~\label{sec:Square-lattice}}

\begin{figure}
\includegraphics[width=1\columnwidth]{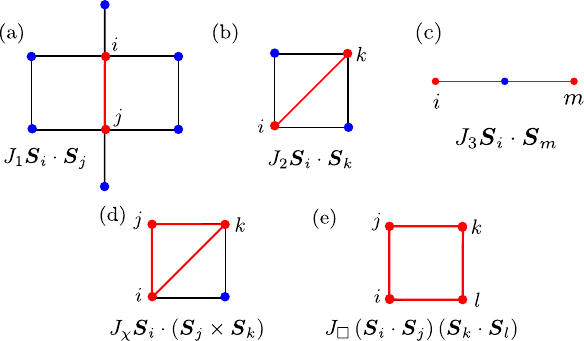}\caption{Representation of the sites involved in fourth-order exchange couplings
on the square lattice. The sites involved in the final expression are indicated in red
while the others are shown in blue (a) $J_{1}$
(b) $J_{2}$ (c) $J_{\chi}$ (d) $J_{\square}$. \label{fig:all-paths-square}}
\end{figure}

The square lattice is more complex than the honeycomb, both due to a larger connectivity, $z =4$ and the possibility of circumscribing a square with four hops.  As such, in addition to $J_2$ and $J_\chi$, we must also consider third neighbor couplings, $J_3$ and ring exchange terms, $J_\square$.  These may all be anisotropic for arbitrary polarizations, and so the general form of the fourth order spin Hamiltonian is,

\begin{align}
H_{{\rm spin}} = &\sum_{\left\langle i,j\right\rangle }J_{1}^{\left(i,j\right)}\boldsymbol{S}_{i}\cdot\boldsymbol{S}_{j}+\sum_{\left\langle \left\langle i,k\right\rangle \right\rangle }J_{2}^{\left(i,k\right)}\boldsymbol{S}_{i}\cdot\boldsymbol{S}_{j}\nonumber \\
 & +\sum_{\triangle}J_{\chi}^{\left(i,j,k\right)}\boldsymbol{S}_{i}\cdot\left(\boldsymbol{S}_{j}\times\boldsymbol{S}_{k}\right)+\!\!\sum_{\left\langle \left\langle \left\langle i,m\right\rangle \right\rangle \right\rangle }\!\!J_{3}^{\left(i,m\right)}\boldsymbol{S}_{i}\cdot\boldsymbol{S}_{m}\nonumber \\
 & +\sum_{\square}\left[J_{\square}^{\left(i,j,k,l\right)}P_{\square}^{\left(i,j,k,l\right)}+J_{\square}^{\left(i,l,j,k\right)}P_{\square}^{\left(i,l,j,k\right)}\right.\nonumber \\
 &\;\; \left.-J_{\square}^{\left(i,k,j,l\right)}P_{\square}^{\left(i,k,j,l\right)}\right]\label{eq:Hamil_square}
\end{align}
with $P_{\square}^{\left(i,j,k,l\right)}$ the product of spins around a plaquette,
\begin{equation}
P_{\square}^{\left(i,j,k,l\right)}=\left(\boldsymbol{S}_{i}\cdot\boldsymbol{S}_{j}\right)\left(\boldsymbol{S}_{k}\cdot\boldsymbol{S}_{l}\right).\label{eq:P_ring}
\end{equation}
The notation of $ij, ijk,$ and $ijkl$ is given in Fig.~\ref{fig:all-paths-square}, which also represents the intermediate sites involved in generating the terms in this Hamiltonian.  The Hamiltonian simplifies greatly for polarization protocols that do not break lattice symmetries, with $J_{1,2,3}^{i,j}$ losing all direction dependence, and 
\begin{equation}
J_{\square}^{\left(i,j,k,l\right)}=J_{\square}^{\left(i,l,j,k\right)}=J_{\square}^{\left(i,k,j,l\right)},
\end{equation}
making the ring exchange terms similarly isotropic.

The expressions for all of the exchange couplings, up to fourth order, are derived similarly to those on the honeycomb lattice, with the expressions given in Appendix~\ref{sec:Expressions}. Most noticeably, the chiral coupling vanishes uniformly, even for the case of circular polarization, as the term is proportional to $\sin\left[m_{2}\left(\beta_{2}-\beta_{1}\right)\right]=\sin\left(m_{2}\frac{\pi}{2}\right)$, while both $\mathcal{L}$ and $\mathcal{B}$ are proportional to $\cos^{2}\left(m_{2}\frac{\pi}{2}\right)$; hence this term vanishes for all $m_2$.  The $\pi/2$ angle between $\boldsymbol{\delta}_1$ and  $\boldsymbol{\delta}_2$ is ultimately responsible for the absence of chiral coupling, and it would return if a next-nearest neighbor $t_2$ or lattice distortions were included.  The fact that $J_{\chi}$ is zero shows that breaking the time-reversal symmetry dynamically is intrinsically distinct from coupling the system to an external magnetic field, where the effects are not so lattice dependent.

\begin{figure}

\includegraphics[width=1\columnwidth]{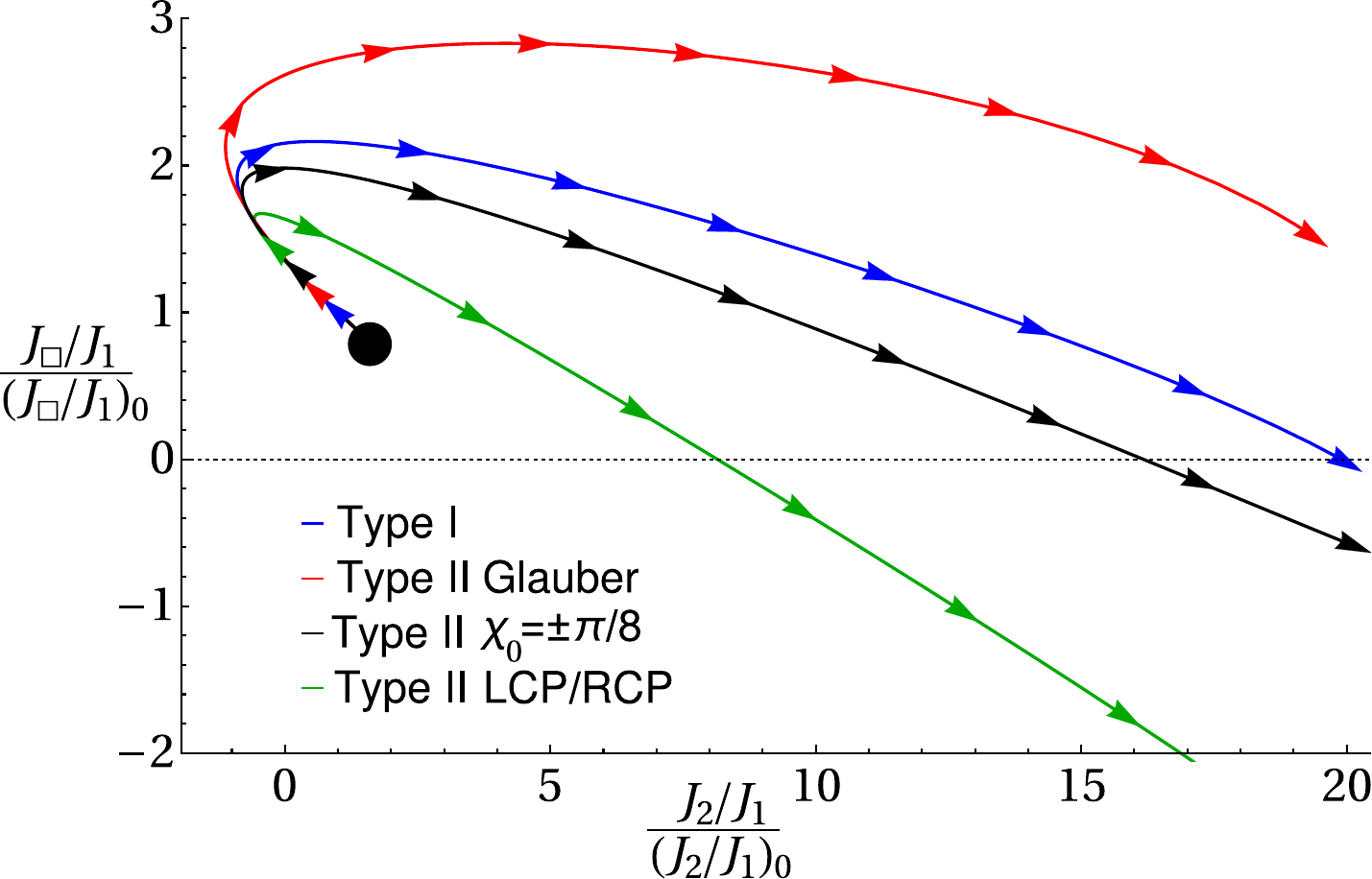}

\caption{Here, we show the region of $J_2/J_1$ and $J_\square/J_1$ phase space on the square lattice that can be accessed by different polarization protocols. The exchange couplings $J_{\square}/J_{1}$ and $J_{2}/J_{1}$ are normalized by their bare values and plotted parametrically as a function of the fluence, $A_0 \in (0,3)$. Different polarization protocols lead to different paths.  Type II Glauber light (red) samples all linearly polarized light equally while type II LCP/RCP (green) samples only the poles of the Poincar\'e sphere. The green curve is identical to circular polarization. The black curve shows a different type II light, sampling the rings $\chi_{0}=\pm \pi/8$, parallel to the equator; all type II light will be bounded by type II Glauber and LCP/RCP. Type I light (blue) samples the Poincar\'e sphere evenly, and is thus a superposition of all different type II light, explaining why it lies within the same fan. Note that the sign of $J_{\square}$ can be tuned by the protocol choice. 
\label{fig:square_lattice_parametric}}
\end{figure}

The nearest-neighbor coupling $J_1$ behaves similarly to the honeycomb lattice, where it is dominated by the second order corrections almost everywhere, but can be driven through zero to become negative.  We again avoid this region in reporting our results, as fourth order perturbation theory is insufficient here.

The square lattice is our first opportunity to examine how different types of polarization can drive materials through distinct regions of phase space.  Here, we examine how $J_2/J_1$ and $J_\square/J_1$ can be tuned parametrically as functions of $A_0$, for the largest allowed $\tilde{t} = 0.0481$ and $\tilde{\Omega} = 2/3$ for different types of unpolarized light, as shown in Fig.~\ref{fig:square_lattice_parametric}.  This figure shows how the two couplings can be enhanced over their bare, time-independent values by type I and several kinds of type II light, with different $\chi_0$'s.  Any type II light may be treated as a superposition of states with different $\chi = \pm\chi_0$'s and will lie in between the two extreme values of $\chi_0 = 0$: type II Glauber light, and type II LCP/RCP light, $\chi_0 = \pi/4$. Aside from the chiral fields, alternating LCP/RCP gives the same results as pure circular polarization. The results for type I light also lies within this fan, as it averages over all $\chi_0$'s equally.  This plot shows how large the enhancement of $J_2/J_1$ really is, with factors of twenty within easy reach, and negative values also possible.  The ring exchange term is harder to enhance, but the sign may be changed, and enhancement factors of $\pm 2$ are possible for various kinds of unpolarized light.

\begin{figure}
\includegraphics[width=1.\columnwidth]{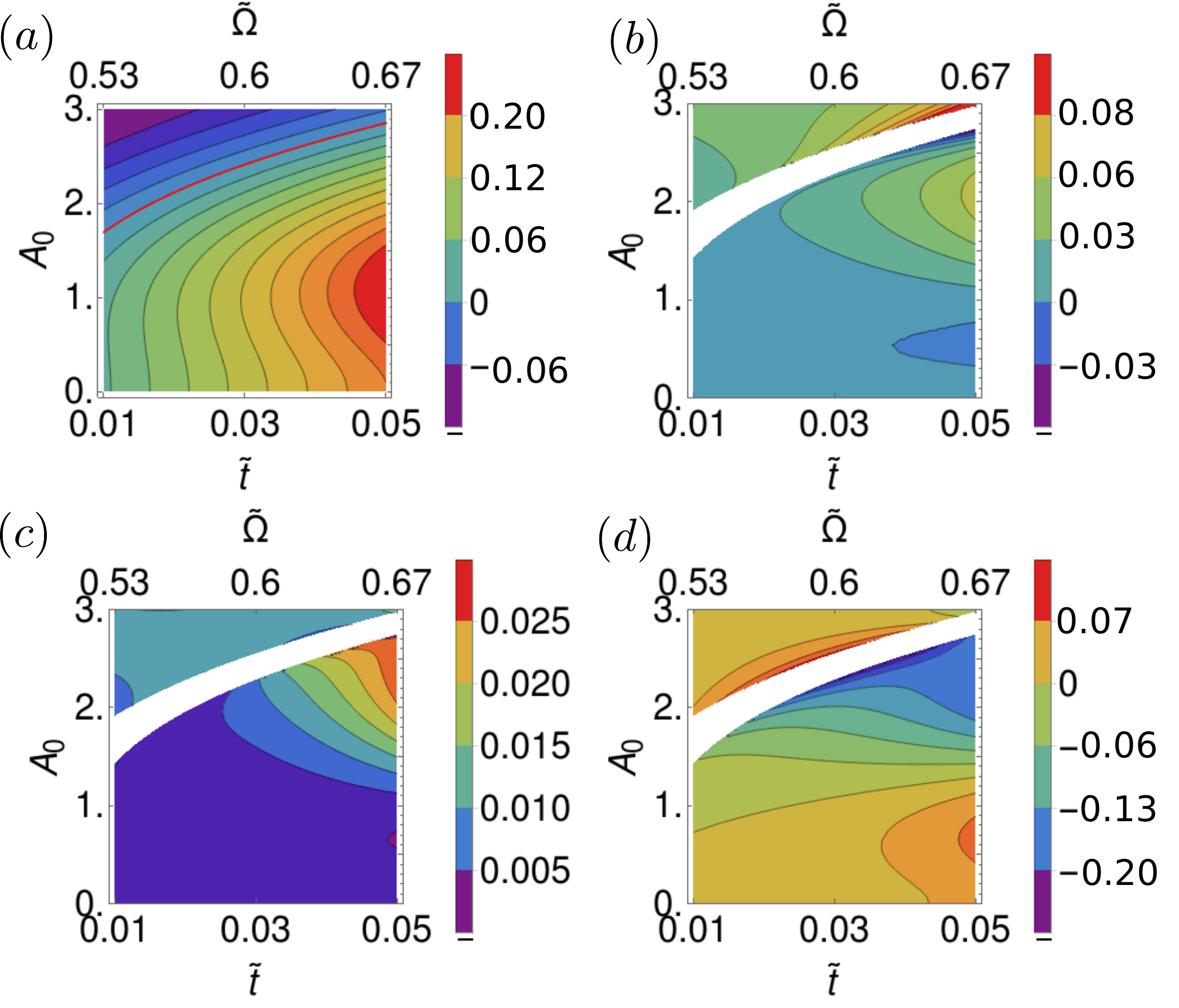}

\caption{Effective couplings on a square lattice, for circular polarized light (a) $J_{1}$ (b) $J_{2}/J_{1}$, (c) $J_{3}/J_{1}$, and (d)
$J_{\square}/J_{1}$, as functions of the dimensionless hopping, $\tilde{t}$ and fluence, $A_0$. The frequency is fixed at $\tilde{\Omega}=1/2+2\sqrt{3}\tilde{t}.$ To fourth order, $J_\chi = 0$. \label{fig:square_lattice_plots}}
\end{figure}

\begin{figure}
\includegraphics[width=1.0\columnwidth]{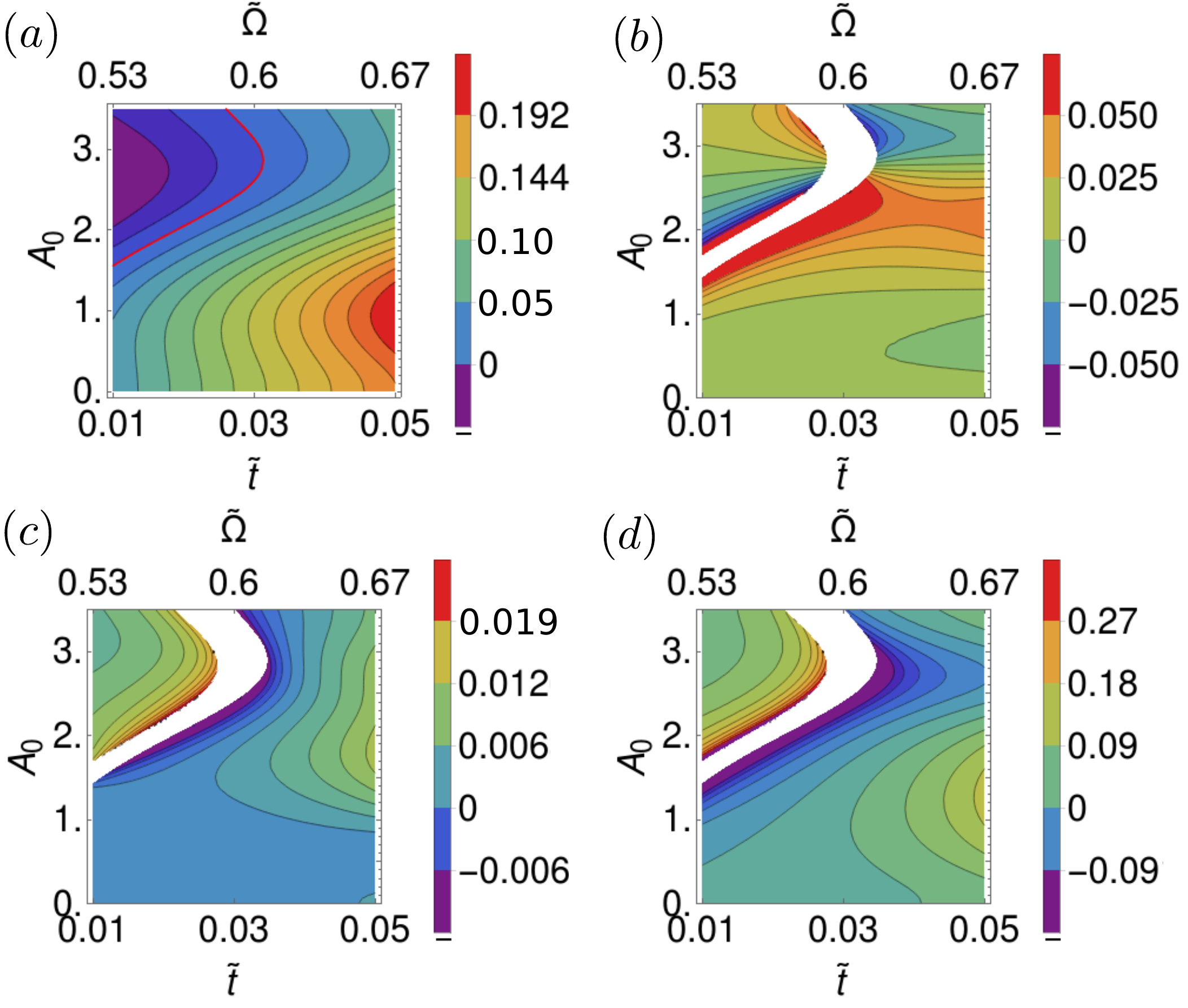}

\caption{Effective couplings on a square lattice for type II Glauber
light as function of $\tilde{t}$ and $A_{0}$, with $\tilde{\Omega}=1/2+2 \sqrt{3} \tilde{t}$ (a) $J_{1}$ (b) $J_{2}/J_{1}$ (c) $J_{3}/J_{1}$,
and (d) $J_{\square}/J_{1}$. \label{fig:square_lattice_plots-unpolarized}}
\end{figure}

In Figs. \ref{fig:square_lattice_plots} and \ref{fig:square_lattice_plots-unpolarized}, we show how all four couplings are tuned as a function of the hopping, $\tilde{t}$ and fluence, $A_0$, with the frequency maximizing proximity to the $\tilde{\Omega} = 1/2$ resonance again chosen: $\tilde{\Omega} = 1/2 + 2\sqrt{3} \tilde{t}$.  We show results for both circular polarization (or LCP/RCP type II light), in Fig.~\ref{fig:square_lattice_plots} and type II Glauber light in Fig.~\ref{fig:square_lattice_plots-unpolarized}, as these are the two extremes that bracket all other kinds of polarized light.

The square lattice is also bipartite, with a stable N\'eel phase that requires $J_2/J_1 \sim 0.4$~\cite{Balents_PRB_2012,Liu_PRB_2018,Mezzacapo_PRB_2012} or $J_\square/J_1 \sim -2$~\cite{Sandvik2007PRL} to destabilize.  Despite the large enhancements, due to the small initial values for allowed $\tilde{t}$'s these regions are unfortunately out of reach for materials that do not already have large $J_2$'s due to other pathways, like next-nearest-neighbor hoppings, $t_2$.  However, about $25\%$ of the critical $J_2/J_1$ can be supplied, so it may be possible to tune materials already close to the transition across the transition at $J_2/J_1 =0.4$, into a quantum disordered regime~\cite{Balents_PRB_2012,Liu_PRB_2018,Mezzacapo_PRB_2012}.

\begin{figure}
\includegraphics[width=1\columnwidth]{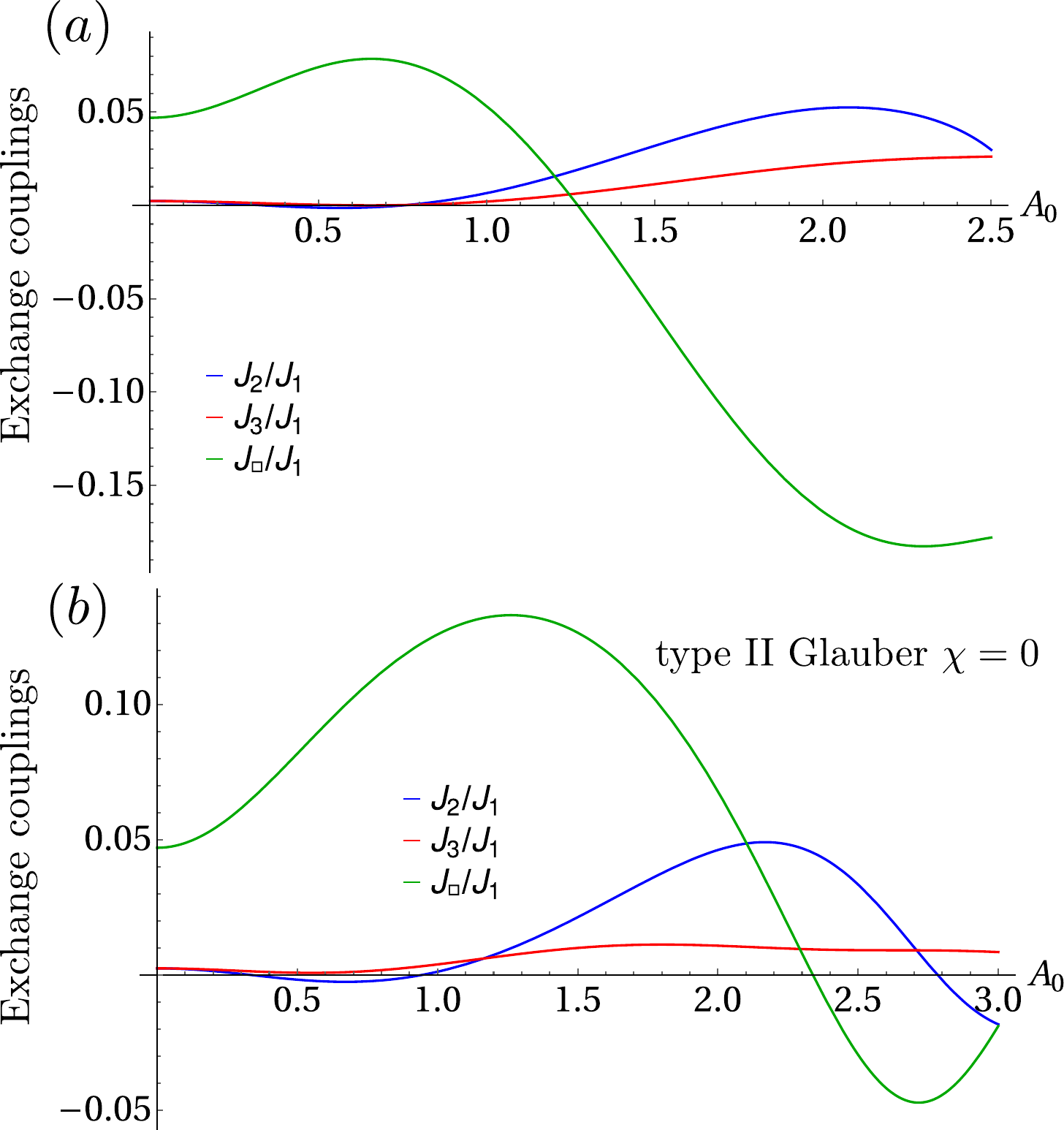}

\caption{Couplings $J_{2}/J_{1}$, $J_{3}/J_{1}$ and $J_{\square}/J_{1}$
on a square lattice as function of $A_{0}$ for fixed $\tilde{t}=\frac{1}{12\sqrt{3}}\approx0.048$
and $\tilde{\Omega}=2/3$. (a) for circularly polarized light (b)
for type II unpolarized light. While typically the unpolarized light
can be used to eliminate the chiral terms, in this particular case,
the chiral couplings are always zero, a consequence of the geometry
of the square lattice. The main distinguishable effect of unpolarized
light is to significantly increase the plaquette term $J_{\square}/J_{1}$.
Given the robustness of the N\'eel phase on the square lattice, these
enhancements are not enough to drive the system to a phase transition.
\label{fig:square_lattice_plots_cut}}
\end{figure}

\subsection{Triangular lattice~\label{sec:Triangular lattice}}

Finally, we come to the triangular lattice, which is the most complicated, but also the most promising.  Here, $z = 6$, leading to a large number of possible paths and possible couplings. As the lattice is non-bipartite, phase transitions and potential spin liquids are more easily accessible.  However, there is a trade-off, as the holon/doublon bandwidth also grows with $z$, and thus the triangular lattice has the lowest maximum $\tilde{t}=0.0372$.  This trade-off means that we end up with similar maximum enhancements of $J_2/J_1$ and other coupling constants, but these enhancements are more effective due to the intrinsic geometric frustration.  The main results on tuning magnetic exchange couplings on the triangular lattice have already been presented in Ref.~\onlinecite{QuitoFlintshort2020}, but here we give more details and can compare to the other two lattices to get a more generic picture. 

\begin{figure}
\includegraphics[width=1\columnwidth]{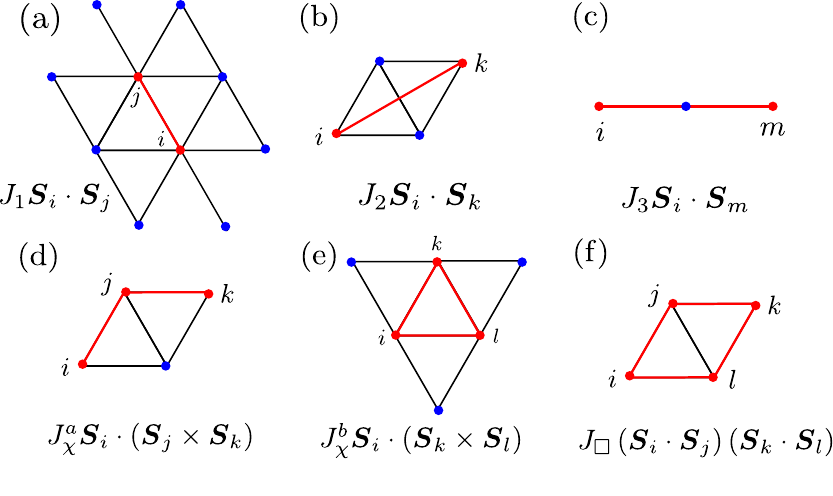}\caption{Representation of the sites involved in fourth order exchange couplings on the triangular lattice, with the sites involved in the final expressions labeled in red. (a) $J_{1}$ (b) $J_{2}$ (c) $J_{3}$ (d) $J_{\chi}^{a}$
(e) $J_{\chi}^{b}$ (f) $J_{\square}$. \label{fig:all-paths-triangular}}
\end{figure}

In this lattice, we find second and third neighbor exchanges, $J_2$, $J_3$, as well as ring exchange around a rhombus, $J_\square$, and two types of chiral fields. One of the chiral fields, involving the isoceles triangle is the same as on the honeycomb lattice, $J_\chi^a$, while the other develops on the equilateral constituent triangles, $J_\chi^b$.  To fourth order, these two are related by $J_\chi^b = -3 J_\chi^a$.  The nearest neighbor coupling $J_1$ behaves similarly to the other two lattices, where it vanishes for a line in the  $\tilde{t}$  and $A_0$ plane that we avoid in our plots. All six effective exchanges in Fig.~\ref{fig:all-paths-triangular}.
The expressions for the triangular couplings are calculated in Appendix \ref{sec:Expressions} in Eqs.~(\ref{eq:J2_generic}),
(\ref{eq:Jsquare_generic}), (\ref{eq:J3_generic}), and (\ref{eq:J_chi_generic})
for $J_{2}$, $J_{\square}$, $J_{3}$ and $J_{\chi}^{a}$, respectively, with  the nearest-neighbor vectors of the triangular lattice
{[}Eq.~(\ref{eq:delta-NN-triangular-honeycomb}){]}.  $J_{1}$ has second order terms from Eqs.~(\ref{eq:J1-order-2}) and fourth order corrections in (\ref{eq:dJ1-triang-CP}) and (\ref{eq:dJ1-triang-LP}) for circular and linear polarizations, respectively.

\begin{figure}
\includegraphics[width=1\columnwidth]{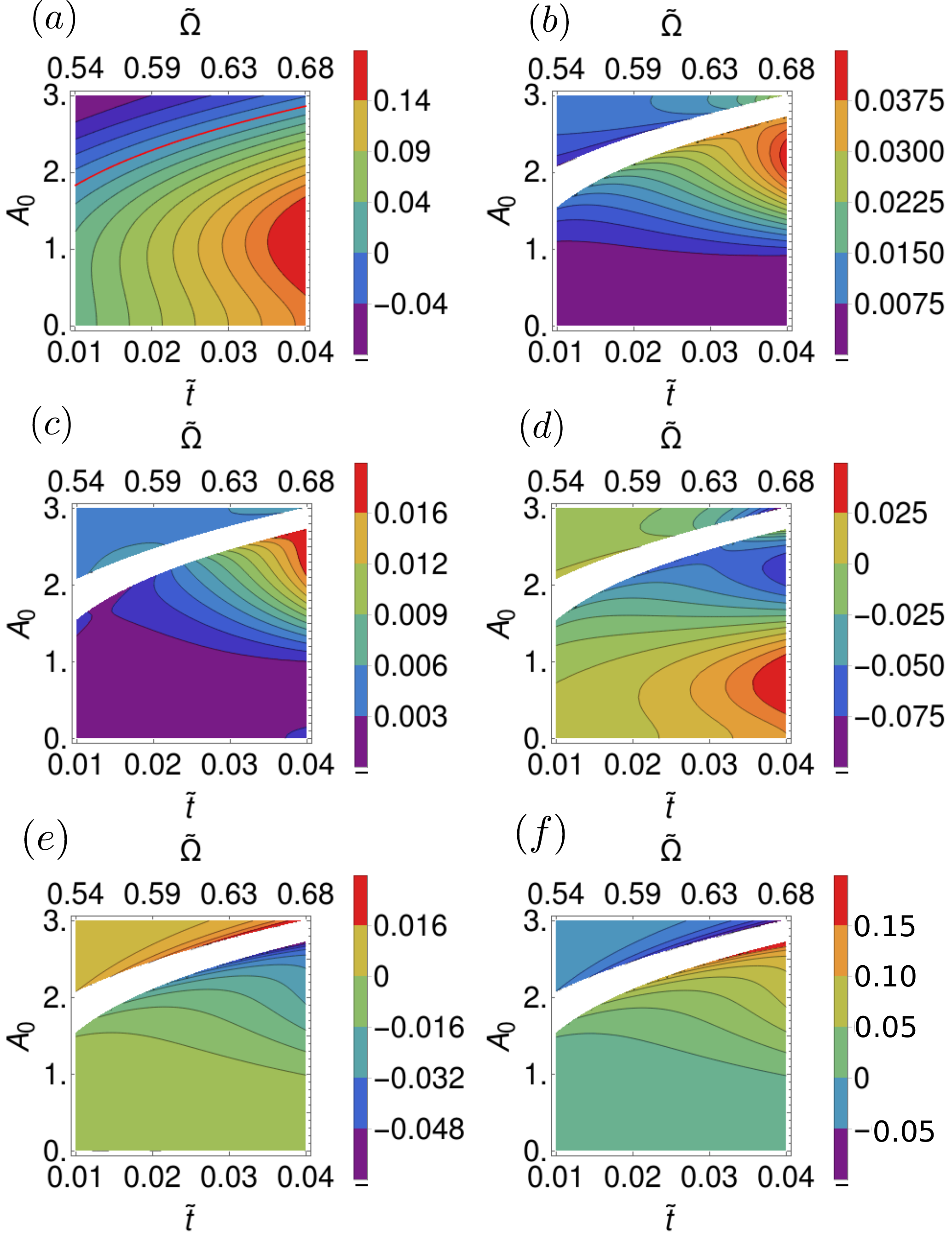}\caption{The exchange couplings on the triangular lattice for circularly polarized
light. The six distinct exchange couplings, following the notation
of Fig.~\ref{fig:all-paths-triangular} are: (a) $J_{1}$ (b) $J_{2}/J_{1}$
(c) $J_{3}/J_{1}$ (d) $J_{\square}/J_{1}$ (e) $J_{\chi}^{a}/J_{1}$
(f) $J_{\chi}^{b}/J_{1}$. The frequency is set to be $\tilde{\Omega}=1/2+2\sqrt{5}\tilde{t}$.
Generically, $J_{\chi}^{b}=-3J_{\chi}^{a}$. \label{fig:triangular_lattice_plots}}
\end{figure}

\begin{figure}
\includegraphics[width=1\columnwidth]{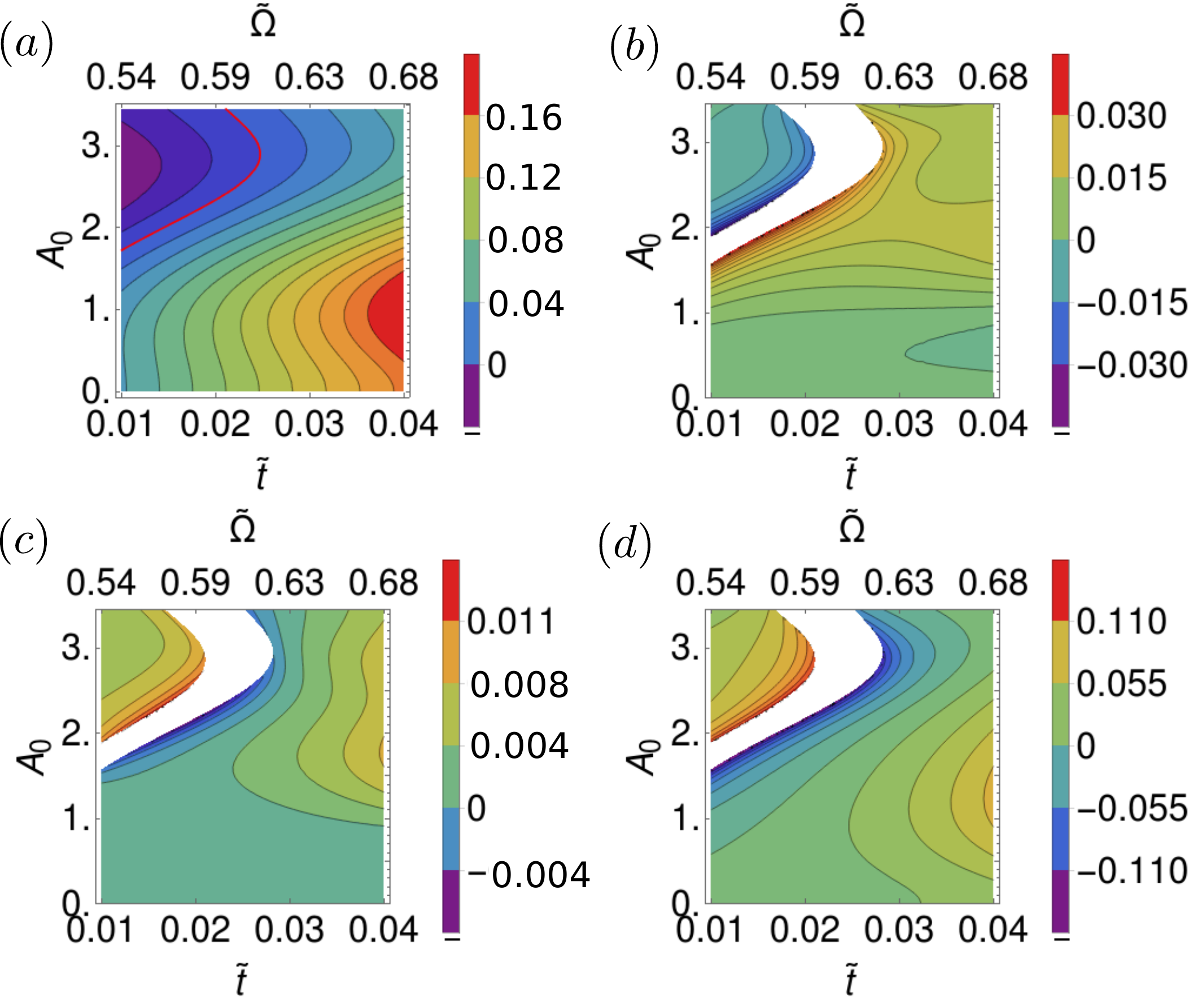}

\caption{The exchange couplings on the triangular lattice for type II Glauber
light as function of the dimensionless hopping, $\tilde{t}$ and dimensionless fluence, $A_0$, with frequency, $\tilde{\Omega}=1/2+2\sqrt{5} \tilde{t}$. (a) $J_{1}$ (b) $J_{2}/J_{1}$ (c) $J_{3}/J_{1}$
(d) $J_{\square}/J_{1}$. The chiral terms, of course, are absent for unpolarized light.
\label{fig:triangular_lattice_plots-unpolarized}}
\end{figure}

As in the honeycomb and square lattices, different types of unpolarized light can drive the exchange couplings through different regions of phase space, but generic unpolarized light is always bracketed by the extremes of type II LCP/RCP, which is equivalent to circular polarization for non-chiral couplings, and type II Glauber.  As such, we show how all six couplings are tuned by the dimensionless hopping, $\tilde{t}$ and fluence, $A_0$ for both circular, Fig.~\ref{fig:triangular_lattice_plots}, and type II Glauber light, Fig.~\ref{fig:triangular_lattice_plots-unpolarized}.  We again choose $\tilde{\Omega} = 1/2+ 2\sqrt{5} \tilde{t}$ to maximize the proximity to the $\tilde{\Omega} = 1/2$ resonance while avoiding heating.  We see that the ratios $J_{2,3}/J_1$ can be massively enhanced by factors of twenty and five, respectively.

There are three potential spin liquids on the $S=1/2$ Heisenberg triangular lattice: a spinon Fermi surface proposed for sufficiently large ring exchange, $J_\square/J_1 \sim 0.18$~\cite{Misguich_PRB_1999,Motrunich_PRB_2005,LeePRL2018}; a Dirac spin liquid accessible by tuning $J_2/J_1 \sim 0.1$~\cite{zhu15,hu14,li15,iqbal16,saadatmand17,wietek17}; and a chiral spin liquid accessible by either strictly tuning $J_\chi/J_1 \sim 0.22$ or $J_2/J_1 \sim 0.08$ and $J_\chi/J_1 \sim 0.03$~\cite{wietek17}.  The spinon Fermi surface state is inaccessible via this kind of Floquet engineering that maxes out $J_\square/J_1 \sim 0.06$ for type II Glauber light; higher values appear in the upper left corner of Fig.~\ref{fig:triangular_lattice_plots-unpolarized}, but these are associated with a ferromagnetic $J_1$.  

The Dirac spin liquid requires a moderate $J_2/J_1 \sim 0.1$, for $J_3 = 0$, with larger $J_3$ pushing the spin liquid boundary out to larger $J_2$~\cite{gong_PRB_2019}.  Circularly polarized or, equivalently, type II LCP/RCP, enhances $J_2/J_1$ the most, up to $\sim 0.04$, while type II Glauber has about half the enhancement.  Some intermediate types of unpolarized light, including type I, enhance $J_2/J_1$ almost as much.

Chiral fields induce a phase transition between the Dirac and chiral spin liquids~\cite{wietek17} if $J_2/J_1 \sim 0.08-0.16$, with  $J_\chi/J_1$ as low as $0.03$.  These numerical phase diagrams are calculating using only $J_\chi^a$, the chiral field for a single plaquette.  We will always also have $J_\chi^b$, which is substantially larger than $J_\chi^a$.  The net flux through a single triangle is likely the relevant quantity, which is $J_{\chi}^{tot}=J_{\chi}^{b}+J_{\chi}^{a}=-2J_{\chi}^{a}$.  This quantity can be as large as $\sim 0.03 J_1$ for the same fluence that maximizes $J_2/J_1$, which raises the possibility of examining this phase transition by tuning the polarization.

The Dirac and chiral spin liquids are not accessible strictly within a single-band Hubbard model, as the absolute value of $J_2/J_1$ is too small.  However, these values can be as large as $\sim 1/3$ of the critical value and so a material with preexisting $J_2$ could potentially have $J_2$ enhanced past the critical value. In this case, both spin liquids would be accessible via unpolarized (Dirac) or circularly polarized (chiral) light.  Most materials with sufficiently low $\tilde{t}$ are actually mediated by superexchange, and not a single band Hubbard model. Tuning the superexchange should be qualitatively similar, but certainly quantitatively different.  In addition, note that theoretical phase diagrams typically only involve two or three couplings, while all couplings shown here are generically present, which will change the phase boundaries.

\section{Conclusions~\label{sec:Conclusions}}

In this paper, we explored how periodic light with different polarization protocols can be used to drive magnetic materials through wide regions of phase space.  We examined the half-filled Hubbard model on the honeycomb, square and triangular lattices to fourth order in perturbation theory, where further neighbor couplings are first generated, and found large enhancements over the equilibrium exchange couplings. We restricted ourselves to polarizations that preserve lattice symmetries, and so considered a range of unpolarized light from type II Glauber light, consisting of all linearly polarized light, to type II LCP/RCP, alternating only left and right circularly polarized light.  These two types of unpolarized light bound the ``tunable'' region of phase space accessible by some non-symmetry breaking polarization protocol, with type I light that samples all polarizations equally, lying in-between the two extremes.  Both quasi-monochromatic type I and type II light are possible to generate experimentally, and so we believe that polarization will prove a key tool in the future to tune strongly correlated materials into interesting regimes out of equilibrium. 

Some interesting questions open for future work involve effects beyond the simple single-band Hubbard model studied here. The effects of phonons and spin-wave interactions, for instance, as a possible source of heating, are left for future investigation. The phonons can also lead to non-trivial changes in hoppings~\cite{ChaudharyRefael_PRR_2020}, which might increase the achievable frustrating exchange couplings. Further work could also explore the effects beyond perturbation theory by numerically simulating the time evolution of the explicit time-dependent interacting Hamiltonian.

\subsection*{Acknowledgments~\label{sec:Acknowledgments}}

We acknowledge helpful discussions with Thomas Iadecola, Peter Orth, Paraj Titum, Thais Trevisan, Chirag Vaswani, and Jigang Wang. V.L.Q and R.F. and were supported by the NSF grant DMR-1555163. RF thanks the Aspen Center for Physics, under the NSF Grant PHY-1607611, for hospitality.

\appendix

\section{Formal structure of the perturbation theory~\label{sec:Formal-structure-pert}}

In this Appendix, we present the formal structure of the perturbation
theory. The main goal is to compute the Floquet-Heisenberg exchange
terms, up to fourth order in $t_{1}$. For that, we take the Brillouin-Wigner
approach~\cite{lindgren1974BW}, as it provides a systematic way
of calculating the perturbation corrections. Another possibility is by Schrieffer-Wolff transformation,~\cite{SchriefferWolfforiginal,PolkovnikovPRL2016} and the results are equivalent for each order. Some aspects shown here were treated compactly in the Supplemental Material of Ref.~\onlinecite{QuitoFlintshort2020}. In this Appendix,  we give further details. 

The first step is to define the states involved. The Hilbert space is enlarged when the Floquet modes are considered, to include a infinite number of copies, labeled by $m$. The identity operator in the enlarged Hilbert space, after combining the Floquet and Fock spaces is 
\begin{align}
\mathds{1} & =\mathds{1}_{\text{Fock}}\otimes \mathds{1}_{\text{\text{Floquet}}}\equiv\mathcal{P}+\mathcal{Q},\label{eq:identity_full}
\end{align}
Here, $\mathcal{P}$ and $\mathcal{Q}$ project onto the ground
and excited state manifolds of the Floquet-Fock Hilbert
space. The resolution of the identity operator in the Fock space $\mathds{1}_{\text{Fock}}$
consists similarly of the projectors $P$ and $Q$
while in the Floquet space, the identity is obtained by summing over
all possible modes. The identities are, therefore,

\begin{align}
\mathds{1}_{\text{Fock}} & =P+Q,\,\,\,\,\,\mathds{1}_{\text{\text{Floquet}}}=\sum_{m=-\infty}^{\infty}P_{F,m}.
\end{align}
Manipulating the resolution of the identity, Eq.~(\ref{eq:identity_full}),
we can separate the ground and excited state manifolds in this enlarged
Floquet-Hilbert space,

\begin{align}
\mathds{1} & =\left(P_{F,0}+\sum_{m\ne0}P_{F,m}\right)\otimes\left(P+Q\right),\nonumber \\
 & =P_{F,0}P+\sum_{m=-\infty}^{+\infty}P_{F,m}Q+\sum_{m\ne0}P_{F,m}P,\\
 & \equiv\mathcal{P}+\mathcal{Q}
\end{align}
The total ground state projector $\mathcal{P}$ has been identified
as the tensor product of the Fock and Floquet ground state manifolds,
$\mathcal{P}=P\otimes P_{F,0}$, while the projector onto excited
manifolds is

\begin{equation}
\mathcal{Q}=\sum_{m=-\infty}^{+\infty}P_{F,m}Q+\sum_{m\ne0}P_{F,m}P.
\end{equation}
The novel effects in the Floquet perturbation
theory comes from the second term of the r.h.s, which projects
onto the fermionic ground state manifold when the system is excited ($m\ne0$) in Floquet
space. From the collection of excited states, it is convenient to
define the so-called resolvent operator, which takes into account the excited states and energy denominators.
Given the two terms of $\mathcal{Q}$, we define the resolvent  $\mathcal{R}=\mathcal{R}_{1}+\mathcal{R}_{2}$, 

\begin{align}
\mathcal{R}_{1} & =\frac{\sum_{m}P_{F,m}Q}{E_{0}-\mathcal{H}_{0}},\label{eq:R_1}\\
\mathcal{R}_{2} & =\frac{\sum_{m\ne0}P_{F,m}P}{E_{0}-\mathcal{H}_{0}}.\label{eq:R_2}
\end{align}
$E_{0}$ is the ground state energy of $\mathcal{H}_{0}$, the time-independent interacting Hamiltonian, with $t_{1}=0$. $E_0$
is zero at half-filling. The definition of $\mathcal{R}$ generically assumes that the non-perturbed Hamiltonian
$H_{0}$ can be exactly solved, and its energies and eigenstates can be used as the building block of the perturbation theory, by the procedure we show next.

The wave operator $\mathcal{W}$, recursively, is defined by ~\cite{lindgren1974BW}

\begin{equation}
\mathcal{W}=\mathcal{P}+\mathcal{R}\left(\mathcal{V}\mathcal{W}-\mathcal{W}\mathcal{V}\mathcal{W}\right).\label{eq:W-eq}
\end{equation}
The low-energy spin Hamiltonian follows from $\mathcal{W}$,

\begin{align}
H_{{\rm spin}}^{\left(m\Omega\lessapprox U\right)} & =\mathcal{P}\mathcal{H}_{0}\mathcal{P}+\mathcal{P}\mathcal{V}\mathcal{W}=\mathcal{P}\mathcal{V}\mathcal{W},\label{eq:effect-Hamilt}
\end{align}
where we use that the projection of $\mathcal{H}_{0}$
onto the ground state is zero. The equation for the wave operator is solved recursively in powers of the perturbation potential $\mathcal{V}$. The order of the expansion
in $\mathcal{W}$ controls the order of the effective Hamiltonian.
Notice that Eq.~(\ref{eq:effect-Hamilt}) has an extra factor of
$\mathcal{V}$ and, therefore, the contributions of order $i$ in
$H_{{\rm spin}}^{\left(\Omega\lessapprox U\right)}$ comes from $\mathcal{W}^{\left(i-1\right)}.$
The zeroth order term from Eq.~(\ref{eq:W-eq}) to $\mathcal{W}$
is~\cite{lindgren1974BW} $\mathcal{W}^{\left(0\right)}=\mathcal{P}=0$,
since $\mathcal{P}$ projects
onto the Fock ground state with one electron per site while $\mathcal{V}$ necessarily creates empty and doubly occupied states. By similar arguments, one can show that all terms with an even number off $\mathcal{V}$ insertions in $\mathcal{W}$ will also vanish. The first and third order contributions to $\mathcal{W}$ are~\cite{lindgren1974BW},

\begin{align}
\mathcal{W}^{\left(1\right)} & =\mathcal{R}\mathcal{V}\mathcal{P},\\
\mathcal{W}^{\left(3\right)} & =\mathcal{R}\mathcal{V}\mathcal{R}\mathcal{V}\mathcal{R}\mathcal{V}\mathcal{P}-\mathcal{R}^{2}\mathcal{V}\mathcal{P}\mathcal{V}\mathcal{R}\mathcal{V}\mathcal{P}.
\end{align}
$\mathcal{W}^{\left(1\right)}$
and $\mathcal{W}^{\left(3\right)}$ lead to the effective Hamiltonian [see Eq.~(\ref{eq:effect-Hamilt})]

\begin{align}
\mathcal{\mathcal{H}}^{\left(2\right)} & =\mathcal{P}\mathcal{V}\mathcal{R}\mathcal{V}\mathcal{P},\label{eq:H-spins-2nd-order}\\
\mathcal{\mathcal{H}}^{\left(4\right)} & =\mathcal{P}\mathcal{V}\mathcal{R}\mathcal{V}\mathcal{R}\mathcal{V}\mathcal{R}\mathcal{V}\mathcal{P}-\left(\mathcal{P}\mathcal{V}\mathcal{R}^{2}\mathcal{V}\mathcal{P}\right)\mathcal{\mathcal{H}}^{\left(2\right)}.\label{eq:H-spins-4th-order}
\end{align}
The perturbation, $\mathcal{V}$, (\ref{eq:V-Floquet}) when projected onto Floquet
spaces yields $P_{F,m_{1}}\mathcal{V}P_{F,m_{2}}=\mathcal{V}_{m_{1}-m_{2}}$.

The second-order couplings $\mathcal{\mathcal{H}}^{\left(2\right)}$
are computed from $\mathcal{R}$ and noticing, from Eq.~(\ref{eq:R_2}), that
$\mathcal{R}_{2}\mathcal{V}\mathcal{P}=0$ since $P\mathcal{V}P=0$. By inserting the resolvent
$\mathcal{R}_{1}$ explicitly into Eq.~(\ref{eq:H-spins-2nd-order}),
we arrive at Eq.~(\ref{eq:second-order-H}). Notice that
$\mathcal{R}_{2}$ does not lead to finite contributions at this particular order. 

The contributions in third-order perturbation theory sum to zero for any fixed polarization. This is a generalization of the  circularly polarized light case studied in Ref~\onlinecite{ClaassenNatComm2017}.  

Now we turn to the fourth order corrections. From the resolvent $\mathcal{R}$
and Eq.~(\ref{eq:H-spins-4th-order}), we find two possible intermediate steps, with either $\mathcal{R}_{1}$
or $\mathcal{R}_{2}$. By splitting the
contributions, we get

\begin{align}
\mathcal{\mathcal{H}}^{\left(4\right)} & =\mathcal{P}\mathcal{V}\mathcal{R}_{1}\mathcal{V}\mathcal{R}_{1}\mathcal{V}\mathcal{R}_{1}\mathcal{V}\mathcal{P}+\mathcal{P}\mathcal{V}\mathcal{R}_{1}\mathcal{V}\mathcal{R}_{2}\mathcal{V}\mathcal{R}_{1}\mathcal{V}\mathcal{P}+\nonumber \\
 & -\left(\mathcal{P}\mathcal{V}\mathcal{R}_{1}^{2}\mathcal{V}\mathcal{P}\right)\mathcal{\mathcal{H}}^{\left(2\right)}.\label{eq:order-4-generic}
\end{align}
The Hilbert space of the problem is again mapped back to the Fock
space of the fermions. By writing $P$ and $Q$ explicitly, we arrive at Eqs.~(\ref{eq:H4a}),
(\ref{eq:H4b}), and (\ref{eq:H4c}) of the main text.

\begin{widetext}

\section{Fourth-order results of the Floquet-Hubbard perturbation theory on
the honeycomb lattice~\label{sec:Floquet-Hubbard-pert}}

The goal in this Appendix is to give further details and some intuition
for how to evaluate the fourth-order corrections given in Eqs.~(\ref{eq:H4a}),
(\ref{eq:H4b}) and (\ref{eq:H4c}). Following the notation of the main text,we call $J^{\left(a,b,c\right)}$ a generic exchange coupling coming from $\mathcal{H}_{a,b,c}$. We will restrict ourselves to
3-site problems, which are enough to calculate the corrections of
$J_{2}$ and $J_{\chi}^{a}$ on the honeycomb lattice. More sites
lead to disconnected terms, which cancel out after the sum over all
sites~\cite{lindgren1974BW}. For other lattices,
it is necessary to consider four or more sites, making the calculations
lengthy. For those, we implemented of Eqs.~(\ref{eq:H4a}), (\ref{eq:H4b})
and (\ref{eq:H4c}) in Mathematica.

To simplify the notation, we label the three sites as 1, 2, and 3. Our strategy is to write the
correction for generic hoppings and then insert the expression for $t_{i,j}^{\left(m\right)}$. We will start from Eqs.~(\ref{eq:H4b})
and (\ref{eq:H4c}). These fourth order terms that be computed
using the product of the strings calculated in second order, Eq.~(\ref{eq:second-order-string}).

We start from $P\mathcal{V}_{-m_{3}}Q\mathcal{V}_{m_{3}-m_{2}}P\mathcal{V}_{m_{2}-m_{1}}Q\mathcal{V}_{m_{1}}P$,
present in $\mathcal{H}_{b}^{\left(4\right)}$, Eq.~(\ref{eq:H4b}).
As we show next, the only effect is to change the nearest-neighbor exchanges. By
inserting, for instance, indices 1 and 2 in both $\mathcal{V}$ terms,
and using Eq.~(\ref{eq:second-order-string}), we find

\begin{align}
P\mathcal{V}_{-m_{3}}^{\left(1,2\right)}Q_{U}\mathcal{V}_{m_{3}-m_{2}}^{\left(1,2\right)}P\mathcal{V}_{m_{2}-m_{1}}^{\left(1,2\right)}Q_{U}\mathcal{V}_{m_{1}}^{\left(1,2\right)}P & =\left(t_{1,2}^{\left(-m_{3}\right)}t_{2,1}^{\left(m_{3}-m_{2}\right)}+t_{2,1}^{\left(-m_{3}\right)}t_{1,2}^{\left(m_{3}-m_{2}\right)}\right)\times\nonumber \\
\times & \left(t_{1,2}^{\left(m_{2}-m_{1}\right)}t_{2,1}^{\left(m_{1}\right)}+t_{2,1}^{\left(m_{2}-m_{1}\right)}t_{1,2}^{\left(m_{1}\right)}\right)\left(2\boldsymbol{S}_{1}\cdot\boldsymbol{S}_{2}-\frac{1}{2}\right)^{2}.
\end{align}
Using the identity for spin 1/2, $\left(\boldsymbol{S}_{a}\cdot\boldsymbol{S}_{b}\right)^{2}=-\frac{1}{2}\boldsymbol{S}_{a}\cdot\boldsymbol{S}_{b}+\frac{3}{16}$,
this term gives, besides a constant, a nearest-neighbor $J_{1}$ exchange between
sites 1 and 2

\begin{align}
J_{1}^{\left(b\right)} & =-2\left(t_{1,2}^{\left(-m_{3}\right)}t_{2,1}^{\left(m_{3}-m_{2}\right)}+t_{2,1}^{\left(-m_{3}\right)}t_{1,2}^{\left(m_{3}-m_{2}\right)}\right)\left(t_{1,2}^{\left(m_{2}-m_{1}\right)}t_{2,1}^{\left(m_{1}\right)}+t_{2,1}^{\left(m_{2}-m_{1}\right)}t_{1,2}^{\left(m_{1}\right)}\right).
\end{align}
Going now to the case where the first hoppings are from sites 1 to
2, while the second ones are from 2 to 3, we obtain, again using Eq.~(\ref{eq:second-order-string}),

\begin{align}
\left[P\mathcal{V}_{-m_{3}}^{\left(1,2\right)}Q_{U}\mathcal{V}_{m_{3}-m_{2}}^{\left(1,2\right)}P\right]\left[P\mathcal{V}_{m_{2}-m_{1}}^{\left(2,3\right)}Q_{U}\mathcal{V}_{m_{1}}^{\left(2,3\right)}P\right] & =\left(t_{1,2}^{\left(-m_{3}\right)}t_{2,1}^{\left(m_{3}-m_{2}\right)}+t_{2,1}^{\left(-m_{3}\right)}t_{1,2}^{\left(m_{3}-m_{2}\right)}\right)\left(2\boldsymbol{S}_{1}\cdot\boldsymbol{S}_{2}-\frac{1}{2}\right)\times\nonumber \\
 & \times\left(t_{3,2}^{\left(m_{2}-m_{1}\right)}t_{3,2}^{\left(m_{1}\right)}+t_{3,2}^{\left(m_{2}-m_{1}\right)}t_{2,3}^{\left(m_{1}\right)}\right)\left(2\boldsymbol{S}_{2}\cdot\boldsymbol{S}_{3}-\frac{1}{2}\right).\label{eq:1,2-3,4}
\end{align}
Chiral term and next-nearest-neighbor exchanges are generated from this term, which
can be seen by using the identity

\begin{equation}
\left(\boldsymbol{S}_{a}\cdot\boldsymbol{S}_{b}\right)\left(\boldsymbol{S}_{b}\cdot\boldsymbol{S}_{c}\right)=-\frac{i}{2}\boldsymbol{S}_{a}\cdot\left(\boldsymbol{S}_{b}\times\boldsymbol{S}_{c}\right)+\frac{1}{4}\boldsymbol{S}_{a}\cdot\boldsymbol{S}_{c}.
\end{equation}
The complete effective couplings are found by summing the contribution
coming from changing $\left(1\rightleftarrows3\right)$ in Eq.~(\ref{eq:1,2-3,4}),

\begin{align}
J_{\chi}^{\left(b\right)} & =-2i\left[\left(t_{1,2}^{\left(-m_{3}\right)}t_{2,1}^{\left(m_{3}-m_{2}\right)}+t_{2,1}^{\left(-m_{3}\right)}t_{1,2}^{\left(m_{3}-m_{2}\right)}\right)\left(t_{3,2}^{\left(m_{2}-m_{1}\right)}t_{3,2}^{\left(m_{1}\right)}+t_{3,2}^{\left(m_{2}-m_{1}\right)}t_{2,3}^{\left(m_{1}\right)}\right)-\left(1\rightleftarrows3\right)\right],\label{eq:Jchib}\\
J_{2}^{\left(b\right)} & =\left(t_{1,2}^{\left(-m_{3}\right)}t_{2,1}^{\left(m_{3}-m_{2}\right)}+t_{2,1}^{\left(-m_{3}\right)}t_{1,2}^{\left(m_{3}-m_{2}\right)}\right)\left(t_{3,2}^{\left(m_{2}-m_{1}\right)}t_{3,2}^{\left(m_{1}\right)}+t_{3,2}^{\left(m_{2}-m_{1}\right)}t_{2,3}^{\left(m_{1}\right)}\right)+\left(1\rightleftarrows3\right),\label{eq:J2b}\\
J_{1}^{\left(b\right)} & =-J_{2}^{\left(b\right)}.\label{eq:J1b}
\end{align}
Notice the relative minus sign in the expression for $J_{\chi}$ coming
from changing the spin operators $\boldsymbol{S}_{1}\cdot\left(\boldsymbol{S}_{2}\times\boldsymbol{S}_{3}\right)=-\boldsymbol{S}_{3}\cdot\left(\boldsymbol{S}_{2}\times\boldsymbol{S}_{1}\right)$.
If the hoppings are real, this minus sign guarantees that the chiral
term vanishes. More generically, a constant overall phase also makes
this term zero. As argued in the main text, this is why the
chiral terms are absent for linear polarization, as expected by symmetry.

The expressions derived so far are generic, and we now use the explicit
form of the hoppings. Following the notation
of the main text, we will use reduced variables $\tilde{\Omega}=\frac{\Omega}{U}$
and $\tilde{t}=\frac{t_{1}}{U}$ and set the overall scale $t_{1}=1$. From
Eqs.~(\ref{eq:Jchib}), (\ref{eq:J2b}) and (\ref{eq:J1b}), the
couplings become 
\begin{align}
J_{\chi}^{\left(b\right)} & =16\mathcal{B}\left(\boldsymbol{m}\right)\sin\left[m_{2}\left(\beta_{1}-\beta_{2}\right)\right],\\
J_{2}^{\left(b\right)} & =8\mathcal{B}\left(\boldsymbol{m}\right)\cos\left[m_{2}\left(\beta_{1}-\beta_{2}\right)\right],\\
J_{1}^{\left(b\right)} & =-J_{2}^{\left(b\right)},
\end{align}
with $\mathcal{B}\left(\boldsymbol{m}\right)$ defined in Eq.~(\ref{eq:B_m}).

The terms of $\mathcal{H}_{c}^{\left(4\right)}$, Eq.~(\ref{eq:H4c}),
follow from the above equations by setting $m_{2}=0$ in the numerator.
After summing over $m_{1}$ and $m_{3}$, the chiral coupling vanishes,
while nearest-neighbor and next-nearest-neighbor contributions are

\begin{align}
J_{2}^{\left(a\right)} & =-4\mathcal{G}\left(\boldsymbol{m}\right),\\
J_{1}^{\left(a\right)} & =-J_{1}^{\left(c\right)},
\end{align}
with $\mathcal{G}\left(\boldsymbol{m}\right)$ defined in Eq.~(\ref{eq:G_m}).

Finally, we calculate the the contributions from Eq.~(\ref{eq:H4a}).
These are purely fourth-order terms that cannot be written by squaring
second-order ones, as done in the previous calculations of this Appendix.
Including the site index, a generic string $\mathcal{S}$ to be calculated
is

\begin{equation}
\mathcal{S}\left(i,\ldots,p\right)=P\mathcal{V}_{-m_{3}}^{\left(i,j\right)}Q\mathcal{V}_{m_{3}-m_{2}}^{\left(k,l\right)}Q\mathcal{V}_{m_{2}-m_{1}}^{\left(m,n\right)}Q\mathcal{V}_{m_{1}}^{\left(o,p\right)}P\label{eq:4th-order-strings}
\end{equation}
Once the path has been chosen, the hoppings that are being multiplied
are completely specified. The remaining task is to decompose the final
operator in terms of nearest-neighbor next-nearest-neighbor and chiral spin terms. This procedure
is straightforward and can be easily implemented in symbolic softwares,
such as Mathematica. Generically, Eq.~(\ref{eq:H-spins-4th-order})
yields

\begin{equation}
\mathcal{S}\left(i,\ldots,p\right)=t_{i,j}^{\left(-m_{3}\right)}t_{k,l}^{\left(m_{3}-m_{2}\right)}t_{m,n}^{\left(m_{2}-m_{1}\right)}t_{o,p}^{\left(m_{1}\right)}\left[\alpha\boldsymbol{S}_{1}\cdot\boldsymbol{S}_{2}+\beta\boldsymbol{S}_{1}\cdot\boldsymbol{S}_{3}+\gamma\boldsymbol{S}_{1}\cdot\left(\boldsymbol{S}_{2}\times\boldsymbol{S}_{3}\right)\right]\label{eq:S_string}
\end{equation}
We list all the non-vanishing contributions in Table~\ref{tab:Fourth-order-contributions}.
To be more concrete, as an example, the third row of Table~\ref{tab:Fourth-order-contributions}
leads to

\begin{equation}
\mathcal{S}\left(1,2,2,3,2,1,3,2\right)=t_{1,2}^{\left(-m_{3}\right)}t_{2,3}^{\left(m_{3}-m_{2}\right)}t_{2,1}^{\left(m_{2}-m_{1}\right)}t_{3,2}^{\left(m_{1}\right)}\left[-1\boldsymbol{S}_{1}\cdot\boldsymbol{S}_{2}+\boldsymbol{S}_{1}\cdot\boldsymbol{S}_{3}-2i\boldsymbol{S}_{1}\cdot\left(\boldsymbol{S}_{2}\times\boldsymbol{S}_{3}\right)\right]
\end{equation}

\begin{table}
\begin{centering}
\begin{tabular}{|c|c|c|c|}
\hline 
path $\left(i,j\right)\rightarrow\text{\ensuremath{\left(k,l\right)\rightarrow\left(m,n\right)\rightarrow\left(o,p\right)}}$ & $\alpha$ & $\beta$ & $\gamma$\tabularnewline
\hline 
\hline 
$\left(1,2\right)\rightarrow\left(2,3\right)\rightarrow\left(3,2\right)\rightarrow\left(2,1\right)$ & -2 & 0 & 0\tabularnewline
\hline 
$\left(2,1\right)\rightarrow\left(3,2\right)\rightarrow\left(2,3\right)\rightarrow\left(1,2\right)$ & -2 & 0 & 0\tabularnewline
\hline 
$\left(1,2\right)\rightarrow\left(2,3\right)\rightarrow\left(2,1\right)\rightarrow\left(3,2\right)$ & -1 & 1 & $-2i$\tabularnewline
\hline 
$\left(2,1\right)\rightarrow\left(3,2\right)\rightarrow\left(1,2\right)\rightarrow\left(2,3\right)$ & -1 & 1 & $-2i$\tabularnewline
\hline 
$\left(2,3\right)\rightarrow\left(1,2\right)\rightarrow\left(3,2\right)\rightarrow\left(2,1\right)$ & -1 & 1 & $2i$\tabularnewline
\hline 
$\left(3,2\right)\rightarrow\left(2,1\right)\rightarrow\left(2,3\right)\rightarrow\left(1,2\right)$ & -1 & 1 & $2i$\tabularnewline
\hline 
\end{tabular}
\par\end{centering}
\caption{Fourth-order contributions of Eq.~(\ref{eq:H4a}) coming from different
paths, using the notation of Eq.~(\ref{eq:4th-order-strings}). The
$\alpha$, $\beta$ and $\gamma$ columns are the pre-factors defined
in Eq.~(\ref{eq:S_string}). \label{tab:Fourth-order-contributions}}
\end{table}
Collecting the contributions listed in Table~\ref{tab:Fourth-order-contributions},
we get

\begin{align}
J_{\chi}^{\left(a\right)} & =-2i\left[t_{1,2}^{\left(-m_{3}\right)}t_{2,3}^{\left(m_{3}-m_{2}\right)}t_{2,1}^{\left(m_{2}-m_{1}\right)}t_{3,2}^{\left(m_{1}\right)}+t_{2,1}^{\left(-m_{3}\right)}t_{3,2}^{\left(m_{3}-m_{2}\right)}t_{1,2}^{\left(m_{2}-m_{1}\right)}t_{2,3}^{\left(m_{1}\right)}-\left(1\rightleftarrows3\right)\right],\\
J_{2}^{\left(a\right)} & =t_{1,2}^{\left(-m_{3}\right)}t_{2,3}^{\left(m_{3}-m_{2}\right)}t_{2,1}^{\left(m_{2}-m_{1}\right)}t_{3,2}^{\left(m_{1}\right)}+t_{2,1}^{\left(-m_{3}\right)}t_{3,2}^{\left(m_{3}-m_{2}\right)}t_{1,2}^{\left(m_{2}-m_{1}\right)}t_{2,3}^{\left(m_{1}\right)}+\left(1\rightleftarrows3\right),\\
J_{1}^{\left(a\right)} & =-J_{2}^{\left(a\right)}-2\left(t_{1,2}^{\left(-m_{3}\right)}t_{2,3}^{\left(m_{3}-m_{2}\right)}t_{3,2}^{\left(m_{2}-m_{1}\right)}t_{2,1}^{\left(m_{1}\right)}+\left(1\rightleftarrows3\right)\right).
\end{align}

We now use the explicit form of the hoppings. Already anticipating
the sums over $m_{1},m_{2},m_{3}$ and the symmetry of the $m_{1}$
and $m_{3}$ indices under summation (see Eq.~(\ref{eq:H4a})), we
change $m_{1}\rightleftarrows m_{3}$ in the last terms. Also restoring
the minus sign of Eq.~(\ref{eq:H4a}), we find

\begin{align}
J_{\chi}^{\left(a\right)} & =-8\mathcal{A}\left(\boldsymbol{m}\right)\sin\left[\left(m_{1}-m_{2}+m_{3}\right)\left(\beta_{1}-\beta_{2}\right)\right],\\
J_{2}^{\left(a\right)} & =-4\mathcal{A}\left(\boldsymbol{m}\right)\cos\left[\left(m_{1}-m_{2}+m_{3}\right)\left(\beta_{1}-\beta_{2}\right)\right],\\
J_{1}^{\left(a\right)} & =-J_{2}^{\left(a\right)}+4\mathcal{A}\left(\boldsymbol{m}\right)\cos\left[\left(m_{1}-m_{3}\right)\left(\beta_{1}-\beta_{2}\right)\right].
\end{align}
with $\mathcal{A}\left(\boldsymbol{m}\right)$ defined in Eq.~(\ref{eq:A_m}).
By combining all the contributions coming to $J_{2}$ and $J_{\chi}$,
we arrive at the Eqs.~(\ref{eq:J2-honey}) and (\ref{eq:Jchi-honey})
of the main text.

\section{Expressions for fourth-order couplings~\label{sec:Expressions}}

In this Appendix, we list the expressions for the magnetic exchange couplings calculated
up to fourth order in perturbation theory, for the three lattices
studied in this work. Following the same order as in the main text,
we start with the honeycomb lattice, then the square and finally the
triangular lattice.

\subsection{Honeycomb lattice}

We start by showing the expressions for the honeycomb lattice. For
arbitrary polarization of light, we find the following fourth-order
contributions to $J_{2}$ and $J_{\chi}$ on the honeycomb lattice

\begin{align}
J_{2}^{\left(i,k\right)} & =\sum_{\boldsymbol{m}}-4\mathcal{A}_{2,1,2,1}\left(\boldsymbol{m}\right)\cos\left[\left(m_{1}-m_{2}+m_{3}\right)\left(\beta_{2}-\beta_{1}\right)\right]+8\mathcal{B}_{1,1,2,2}\left(\boldsymbol{m}\right)\cos\left[m_{2}\left(\beta_{2}-\beta_{1}\right)\right]+4\mathcal{G}_{2,1}\left(\boldsymbol{m}\right),\label{eq:J2-honey}\\
J_{\chi}^{\left(i,j,k\right)} & =\sum_{\boldsymbol{m}}-8\mathcal{A}_{2,1,2,1}\left(\boldsymbol{m}\right)\sin\left[\left(m_{1}-m_{2}+m_{3}\right)\left(\beta_{2}-\beta_{1}\right)\right]-16\mathcal{B}_{1,1,2,2}\left(\boldsymbol{m}\right)\sin\left[m_{2}\left(\beta_{2}-\beta_{1}\right)\right].\label{eq:Jchi-honey}
\end{align}

Now, the corrections for the nearest-neighbor coupling $J_{1}$. We will show the
expressions for linear and circular polarizations, as the expressions
for arbitrary polarization become quite lengthy and were treated in Mathematica. For circular polarization,
defining

\begin{align}
f_{\hexagon}^{\left(CP\right)}\left(\boldsymbol{m}\right) & =\cos\left[\frac{1}{3}\pi\left(m_{1}-m_{3}\right)\right]+(-1)^{m_{1}+m_{3}}\cos\left[\frac{2}{3}\pi\left(m_{1}-m_{3}\right)\right]+\cos\left[\frac{1}{3}\pi\left(m_{1}-m_{2}+m_{3}\right)\right]+\nonumber \\
+ & (-1)^{m_{1}+m_{2}+m_{3}}\cos\left[\frac{2}{3}\pi\left(m_{1}-m_{2}+m_{3}\right)\right],
\end{align}
we find

\begin{align}
\delta J_{1}^{\left(4\right)} & =8f_{\hexagon}^{\left(CP\right)}\left(\boldsymbol{m}\right)\mathcal{A}\left(\boldsymbol{m}\right)+16\left[\cos\left(\frac{\pi m_{2}}{3}\right)+\cos\left(\frac{2\pi m_{2}}{3}\right)+1\right]\mathcal{B}\left(\boldsymbol{m}\right)-24\mathcal{G}\left(\boldsymbol{m}\right).\,\,\,\text{(CP)}\label{eq:J1-honey}
\end{align}
The fourth-order corrections (\ref{eq:J1-honey}) are added to the second order terms from Eq.~(\ref{eq:J1-order-2})
to yield the complete expression for $J_{1}$. As for linear polarization,
the correction of a bond along the $\boldsymbol{\delta}_{1}$ direction
is

\begin{align}
\delta J_{1}^{\left(4\right)} & =8\left[\mathcal{A}_{1,2,2,1}+\mathcal{A}_{2,1,2,1}+(-1)^{m_{1}+m_{3}}\mathcal{A}_{1,2,2,1}+(-1)^{m_{1}+m_{2}+m_{3}}\mathcal{A}_{1,3,3,1}\right]+\nonumber \\
 & +16\left(\mathcal{B}_{1,2}+\mathcal{B}_{1,1}+\mathcal{B}_{3,1}\right)-8\left(\mathcal{G}_{2,1}+\mathcal{G}_{1,1}+\mathcal{G}_{3,1}\right).\,\,\,\text{(LP)}
\end{align}
Here, each of these functions depends on $\phi_l$ for different intermediate links, as indicated in Eq.~(\ref{eq:H4a}) etc.
The correction of bonds along other directions are found by permutation
of the indices of $\mathcal{A}$, $\mathcal{B}$ and $\mathcal{G}$.

\subsection{Square lattice}

The square lattice has more couplings than the honeycomb case. The reader is invited to revisit Fig.~\ref{fig:all-paths-square}
of the main text for the definition of the couplings. The next-nearest-neighbor coupling $J_{2}$
{[}Fig.~\ref{fig:all-paths-square}(b){]} is

\begin{align}
J_{2}^{\left(i,k\right)} & =\sum_{\boldsymbol{m}}-8\left\{ \mathcal{A}_{1,2,2,1}\left(\boldsymbol{m}\right)\cos^{2}\left[\left(m_{1}+m_{3}\right)\frac{\pi}{2}\right]\cos\left[\left(\beta_{2}-\beta_{1}\right)\left(m_{1}-m_{3}\right)\right]+\mathcal{A}_{1,2,1,2}\left(\boldsymbol{m}\right)\cos^{2}\left[\left(m_{1}+m_{2}+m_{3}\right)\frac{\pi}{2}\right]\right.\times\nonumber \\
\times & \left.\cos\left[\left(m_{1}-m_{2}+m_{3}\right)\left(\beta_{2}-\beta_{1}\right)\right]\right\} +8\mathcal{L}_{2,2,1,1}\left(\boldsymbol{m}\right)\cos\left[m_{2}\left(\beta_{2}-\beta_{1}\right)\right]-16\mathcal{B}_{2,1}\left(\boldsymbol{m}\right)\cos\left[\left(\beta_{2}-\beta_{1}\right)m_{2}\right]+8\mathcal{G}_{2,1}\left(\boldsymbol{m}\right),\label{eq:J2_generic}
\end{align}
while the plaquette terms {[}Fig.~\ref{fig:all-paths-square}(e){]}
are

\begin{align}
J_{\square}^{\left(i,j,k,l\right)}= & \sum_{\boldsymbol{m}}32\left\{ \mathcal{A}_{1,2,2,1}\left(\boldsymbol{m}\right)\cos^{2}\left[\left(m_{1}+m_{3}\right)\frac{\pi}{2}\right]\cos\left[\left(\beta_{2}-\beta_{1}\right)\left(m_{1}-m_{3}\right)\right]+\mathcal{A}_{1,2,1,2}\left(\boldsymbol{m}\right)\cos^{2}\left[\left(m_{1}+m_{2}+m_{3}\right)\frac{\pi}{2}\right]\right.\times\nonumber \\
\times & \left.\cos\left[\left(m_{1}-m_{2}+m_{3}\right)\left(\beta_{2}-\beta_{1}\right)\right]\right\} +32\cos\left[m_{2}\left(\beta_{2}-\beta_{1}\right)\right]\mathcal{L}_{2,2,1,1}\left(\boldsymbol{m}\right).\label{eq:Jsquare_generic}
\end{align}
The $J_{3}$ coupling {[}Fig.~\ref{fig:all-paths-square}(c){]} is

\begin{align}
J_{3}^{\left(i,l,m\right)}= & \sum_{\boldsymbol{m}}-4\mathcal{A}_{2,2,2,2}\left(\boldsymbol{m}\right)+8\mathcal{B}_{2,2}\left(\boldsymbol{m}\right)+4\mathcal{G}_{2,2}\left(\boldsymbol{m}\right).\label{eq:J3_generic}
\end{align}
The expressions for $J_2$, $J_{3}$ and $J_\square$ here are the same for both the triangular and square lattices, as the hoppings are topologically identical, as seen by comparing Figs. ~\ref{fig:all-paths-square}(b,c,e) and ~\ref{fig:all-paths-triangular}(b,c,f).  Only $J_3$ will actually be identical though, as it involves only straight line hopping in both cases, while the $J_2$ and $J_\square$ expressions will differ due to the bond angle differences.

The chiral term reads {[}Fig.~\ref{fig:all-paths-square}(d){]}

\begin{equation}
J_{\chi}^{a\left(i,j,k\right)}=\sum_{\boldsymbol{m}}16\left[\mathcal{L}_{2,2,1,1}\left(\boldsymbol{m}\right)+\mathcal{B}_{2,1}\left(\boldsymbol{m}\right)\right]\sin\left[m_{2}\left(\beta_{2}-\beta_{1}\right)\right].\label{eq:J_chi_generic}
\end{equation}
This term of course vanishes on the square lattice as the $\beta_i$ are right angles.

As for the fourth-order correction to $J_{1}$ on the square lattice
(Fig.~\ref{fig:all-paths-square}(a)), we again show only the cases
of circularly and linearly polarized light. For circularly polarized
light,

\begin{align}
\delta J_{1}^{\left(4\right)}= & 16\mathcal{A}\left(\boldsymbol{m}\right)\left\{ 1+\cos^{2}\left[\left(m_{1}+m_{2}+m_{3}\right)\frac{\pi}{2}\right]\cos\left[\left(m_{1}-m_{2}+m_{3}\right)\frac{\pi}{2}\right]+\cos\left[\left(m_{1}-m_{3}\right)\frac{\pi}{2}\right]\cos^{2}\left[\left(m_{1}+m_{3}\right)\frac{\pi}{2}\right]\right\} -\nonumber \\
 & -16\mathcal{L}\left(\boldsymbol{m}\right)+64\mathcal{B}\left(\boldsymbol{m}\right)\cos^{2}\left(\frac{\pi m_{2}}{4}\right)-32\mathcal{G}\left(\boldsymbol{m}\right),\,\,\,\left(\text{CP, square}\right)\label{eq:dJ1-square-CP}
\end{align}
while for linearly polarized light, assuming a vertical bond (along
the $\boldsymbol{\delta}_{1}$ direction), the correction is

\begin{align}
\delta J_{1}^{\left(4\right)} & =\sum_{\boldsymbol{m}}16\left(\cos^{2}\left[\left(m_{1}+m_{3}\right)\frac{\pi}{2}\right]\mathcal{A}_{1,2,1,2}+\cos^{2}\left[\left(m_{1}+m_{2}+m_{3}\right)\frac{\pi}{2}\right]\mathcal{A}_{2,1,1,2}+\mathcal{A}_{1,1,1,1}\right)-\nonumber \\
 & -16\mathcal{L}_{1,1,2,2}+32\left(\mathcal{B}_{1,2}+\mathcal{B}_{1,1}\right)-16\left(\mathcal{G}_{1,2}+\mathcal{G}_{2,1}\right).\,\,\,\left(\text{LP, square}\right)\label{eq:dJ1-square-LP}
\end{align}
Again, the $\phi_l$ dependence is hidden in the $A_l$ dependences of the functions.  As a sanity check, it is easy to verify that in the time-independent, $A_{0}\rightarrow0$
limit, $\delta J_{1}^{\left(4\right)}\rightarrow-24t_{1}^{4}/U^{3}$.

\subsection{Triangular lattice}

For the notation of this Subsection, we refer to Fig.~\ref{fig:all-paths-triangular}.
The results for $J_{2}^{\left(i,k\right)}$, $J_{\square}^{\left(i,j,k,l\right)}$, $J_{3}^{\left(i,l,m\right)}$, and $J_{\chi}^{a\left(i,l,m\right)}$ follow the identical expressions as listed
for the square lattice, with the important difference that the angles
$\beta_{1}$ and $\beta_{2}$ take different values in each lattice
(see Eqs.~(\ref{eq:cos-bl}) and (\ref{eq:sin-bl}) for the definition
of $\beta_{l}$). The chiral term $J_\chi^a$ is given already in Eq.~(\ref{eq:J_chi_generic}), and $J_\chi^b = -3J_\chi^a$

The only truly different forms come with the fourth-order corrections for $J_{1}$, as different numbers of intermediate sites are involved for different lattices. For circular
polarization, $\delta J_{1}^{\left(4\right)}$ is

\begin{align}
\delta J_{1}^{\left(4\right)} & =\sum_{\boldsymbol{m}}8\mathcal{A}\left(\boldsymbol{m}\right)f_{\triangle}^{\left(CP\right)}\left(\boldsymbol{m}\right)-8\mathcal{L}\left(\boldsymbol{m}\right)\left[\cos\left(\frac{\pi m_{2}}{3}\right)+2\cos\left(\frac{2\pi m_{2}}{3}\right)\right]\nonumber \\
 & -16\mathcal{B}\left(\boldsymbol{m}\right)\left[\cos\left(\frac{\pi m_{2}}{3}\right)+2\cos\left(\frac{2\pi m_{2}}{3}\right)+2\right]-40\mathcal{G}\left(\boldsymbol{m}\right),\,\,\,\left(\text{CP, triangular}\right)\label{eq:dJ1-triang-CP}
\end{align}
with $f_{\triangle}^{\left(CP\right)}\left(\boldsymbol{m}\right)$ defined as 

\begin{align}
f_{\triangle}^{\left(CP\right)}\left(\boldsymbol{m}\right) & =2+\cos^{2}\left[\left(m_{1}+m_{3}\right)\frac{\pi}{2}\right]\left(\cos\left[\frac{1}{3}\pi\left(m_{1}-m_{3}\right)\right]+2\cos\left[\frac{2}{3}\pi\left(m_{1}-m_{3}\right)\right]\right)\nonumber \\
 & +\cos^{2}\left[\left(m_{1}+m_{2}+m_{3}\right)\frac{\pi}{2}\right]\left(\cos\left[\frac{1}{3}\pi\left(m_{1}-m_{2}+m_{3}\right)\right]+2\cos\left[\frac{2}{3}\pi\left(m_{1}-m_{2}+m_{3}\right)\right]\right).
\end{align}

The fourth-order correction $\delta J_{1}^{\left(4\right)}$ to a bond along the $\boldsymbol{\delta}_{3}$ direction for a system coupled to linearly polarized light is

\begin{align}
\delta J_{1}^{\left(4\right)} & =\sum_{\boldsymbol{m}}8\left\{ \cos^{2}\left[\left(m_{1}+m_{2}+m_{3}\right)\frac{\pi}{2}\right]+\cos^{2}\left[\left(m_{1}+m_{3}\right)\frac{\pi}{2}\right]\right\} \left(\mathcal{A}_{3,2,3,2}+\mathcal{A}_{2,3,3,2}+\mathcal{A}_{3,1,3,1}+\mathcal{A}_{1,3,3,1}\right)\nonumber \\
 & -8\cos^{2}\left[\left(m_{1}+m_{3}\right)\frac{\pi}{2}\right]\left(\mathcal{A}_{1,2,1,2}+\mathcal{A}_{2,1,1,2}\right)16\mathcal{A}_{2,2,2,2}-8\left(\mathcal{L}_{2,2,3,3}+\mathcal{L}_{3,3,2,2}-\mathcal{L}_{2,2,1,1}\right)\nonumber \\
 & -16\left[2\left(\mathcal{B}_{3,2}+\mathcal{B}_{2,3}\right)+2\mathcal{B}_{3,3}-\mathcal{B}_{1,2}\right]-8\left(2\mathcal{G}_{3,2}+2\mathcal{G}_{2,3}+2\mathcal{G}_{3,3}-\mathcal{G}_{1,2}\right),\,\,\,\left(\text{LP, triangular}\right)\label{eq:dJ1-triang-LP}
\end{align}
Notice that $\delta J_{1}^{\left(4\right)}\rightarrow-28t_{1}^{4}/U^{3}$
as $A_{0}\rightarrow0$. Again, corrections in other directions
are calculating by a proper change of sub-indices.

\end{widetext}

\section{Comparing the second and fourth order contributions to $J_{1}$~\label{sec:J1-order2-order4}}

For most of the regimes we are interested in, the nearest-neighbor coupling, $J_1$ is dominated by the second order contributions, indicating that the perturbation expansion is likely sound.  However, these second order terms can actually drive $J_1$ negative, and near where $J_1$ crosses zero, higher order terms in perturbation theory are likely required, at least for $J_1$.  In this Appendix, we show that, except extremely close to the zero crossing, the second order terms are much larger than the fourth order terms and the perturbation expansion reasonable.

\begin{figure}
\includegraphics[width=1\columnwidth]{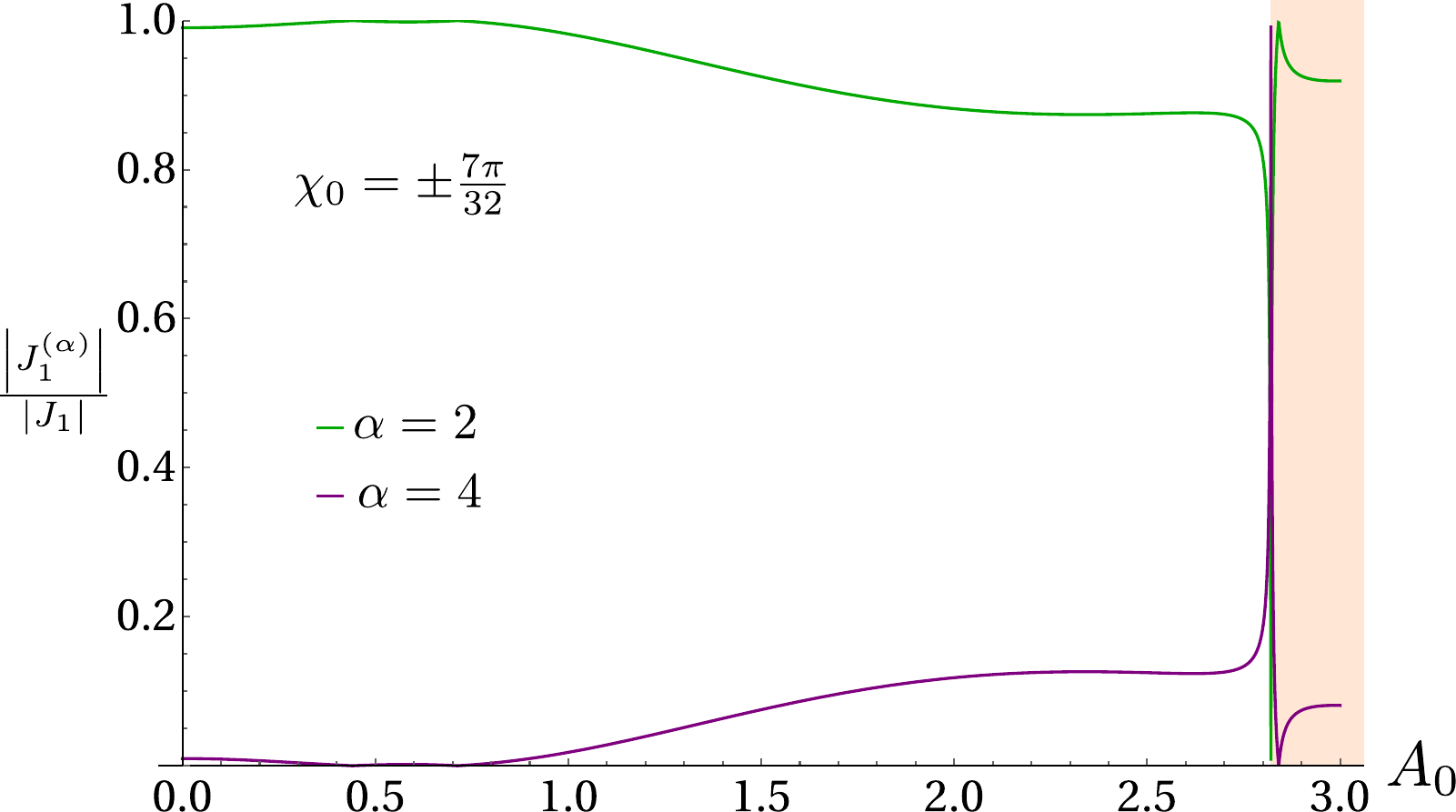}
\caption{The fraction of the two contributions for $J_{1}$, in second (green) and fourth order (purple) perturbation theory on a triangular lattice for type II $\chi_{0}=\pm7\pi/32$ light, with $\tilde{t}=0.037$ and $\Omega=2/3$. The corrections coming from fourth order change $J_{1}$ only by a small fraction
in most regions. Close to $A_{0}=2.8$, $J_{1}$ vanishes up to fourth order and becomes negative for larger values of $A_{0}$ (light orange region).   \label{fig:honeycomb_lattice_J1}}
\end{figure}

We  write 
\begin{align}
J_{1}=J_{1}^{\left(2\right)}+J_{1}^{\left(4\right)}
\end{align}
and plot the results for the triangular lattice in  Fig.~\ref{fig:honeycomb_lattice_J1}. We show the fraction of the two contributions to $J_{1}$, $\left|J_{1}^{\left(2\right)}\right|/\left|J_{1}\right|$ and $\left|J_{1}^{\left(4\right)}\right|/\left|J_{1}\right|$. We keep the maximum fluence $A_{0}$ in the region such that the second-order term yields a substantial fraction, in this case, more than $80\%$ of the total contribution, justifying the perturbative expansion.  This region is contained within the $|J_1| < .01$ region excluded from our plots.

\bibliographystyle{apsrev4-1}
\bibliography{FSLs_references.bib,biblio_Sahoo.bib}

\end{document}